\documentclass[aps,prb,showpacs,notitlepage,reprint, floatfix]{revtex4-1}
\pdfoutput=1
\usepackage{color}
\usepackage{graphicx}
\usepackage{natbib}
\usepackage{braket}
\usepackage{bm}
\usepackage{amsmath}
\usepackage{amssymb}

\begin{document}
\title{Layer-dependent anisotropic electronic structure of freestanding quasi-two-dimensional MoS$_2$}
\author{Jinhua Hong$^{1}$}
\author{Kun Li$^{2}$}
\author{Chuanhong Jin$^{1}$}
\email[Corresponding authors's e-mail addresses: ]{chhjin@zju.edu.cn, xixing.zhang@kaust.edu.sa or jun.yuan@york.ac.uk}
\author{Xixiang Zhang$^3$}
\email[Corresponding authors's e-mail addresses: ]{chhjin@zju.edu.cn, xixing.zhang@kaust.edu.sa or jun.yuan@york.ac.uk}
\author{Ze Zhang$^1$}
\author{Jun Yuan$^{1,4}$}
\email[Corresponding authors's e-mail addresses: ]{chhjin@zju.edu.cn, xixing.zhang@kaust.edu.sa or jun.yuan@york.ac.uk}
\affiliation{$^1$State Key Laboratory of Silicon Materials and School of Materials Science and Engineering, Zhejiang University, Hangzhou, Zhejiang 310027, People's Republic of China}
\affiliation{$^2$Advanced Nanofabrication, Imaging and Characterization Core Lab, King Abdullah University of Science and Technology (KAUST), Thuwal 239955, Kingdom of Saudi Arabia}
\affiliation{$^3$Division of Physical Science and Engineering, King Abdullah University of Science and Technology (KAUST), Thuwal 239955, Kingdom of Saudi Arabia}
\affiliation{$^4$Department of Physics, University of York, Heslington, York, YO10 5DD, UK}

\pacs{42.50.Tx, 42.50.Wk, 37.10.Vz}

%
\begin{abstract}
The anisotropy of the electronic transition is a well-known characteristic of low-dimensional transition-metal dichalcogenides, but their layer-thickness dependence has not been properly investigated experimentally until now. Yet, it not only determines the optical properties of these low-dimensional materials, but also holds the key in revealing the underlying character of the electronic states involved. Here we used both angle-resolved electron energy-loss spectroscopy and spectral analysis of angle-integrated spectra to study the evolution of the anisotropic electronic transition involving the low energy valence electrons in the freestanding MoS$_2$ layers with different thicknesses. We are able to demonstrate that the well-known direct gap at 1.8 eV is only excited by the in-plane polarized field while the out-of-plane polarized optical gap is 2.4$\pm$0.2 eV in monolayer MoS$_2$. This contrasts with the much smaller anisotropic response found for the indirect gap in the few-layer MoS$_2$ systems. In addition, we determined that the joint density of states associated with the indirect gap transition in the multilayer systems and the corresponding indirect transition in the monolayer case has a characteristic three-dimensional-like character. We attribute this to the soft-edge behavior of the confining potential and it is an important factor when considering the dynamical screening of the electric field at the relevant excitation energies. Our result provides a logical explanation, for the large sensitivity of the indirect transition to thickness variation compared with that for the direct transition, in terms of quantum confinement.
\end{abstract}

\maketitle
\section{INTRODUCTION}
Atomically thin molybdenum disulfide (MoS$_2$), as a representative member of the emerging two-dimensional (2D) transition metal dichalcogenides (TMDs) \cite{Wang2012, Ataca2012}, has attracted intensive research efforts owing to its unique structure as well as its novel applications in optoelectronics \cite{Zhang2012, Radisavljevic2011, Radisavljevic2011a, Baugher2013, Yin2012, Lee2012} and valleytronics \cite{Zeng2012, Mak2012}. It shows a strong layer-dependent electronic structure which changes dramatically at the atomically thin limit. For example, the theoretically predicted transition from an indirect to a direct energy gap \cite{Li2007a, Lebegue2009} has been confirmed initially indirectly by the observation of strong photoluminescence (PL) enhancement in the monolayer \cite{Mak2010, Splendiani2010} and now directly by angle-resolved photoemission spectroscopy (ARPES) \cite{Jin2013, Zhang2014}. The observation of strong chiral pumping effect in the band-gap absorption \cite{Zeng2012, Mak2012} has opened up the possibility of dynamical control of valley-specific carrier density in the monolayer system which lacks inversion symmetry. The realization of vertically stacked heterostructures \cite{Geim2013, Georgiou2013} also opens the door to explore new physics and applications through combinations of different atomically thin layers, such as significant extrinsic photoconversion in graphene/TMD/graphene trilayer heterojunction \cite{Britnell2013}.

Despite intense efforts on these novel TMDs, the anisotropic properties of the optical response of these low-dimensional semiconductors, particularly their evolution with the layer thickness, has received much less experimental attention. This is particularly glaring as optical response of bulk MoS$_2$ itself is known to be already highly anisotropic \cite{Liang1973}. Anisotropic electronic excitation plays an important role because of the existence of significant out-of-plane bonding between sulfur and Mo atoms. On the practical level, the out-of-plane optical response is difficult to measure even for bulk MoS$_2$ and related materials as either thick samples with optical-quality surfaces or large-area layered materials are required for experiments with oblique illumination  \cite{Liang1973}. Nevertheless, such measurements are extremely useful to understand the complex electronic structure of atomically-thin TMDs. For example, it can reveal directly the out-of-plane energy gap which is important for the vertical transport in layer-stacking heterostructures \cite{Geim2013} and also complement the partial picture of the electronic structure given by optical absorption \cite{Mak2010, Splendiani2010} which usually probes more efficiently the states sensitive to in-plane electric field of normal incident light. Here we want to emphasize that the periodic band-structure of the bulk MoS$_2$ system will evolve into the "discrete atomic levels" in the out-of-plane direction when the thickness is reduced to atomically thin MoS$_2$ (0.6-3 nm thickness). For consistence, we will use the term "energy gap" or "optical gap" transitions for the out-of-plane excitation in single or few-layer MoS$_2$ systems in order to make the correspondence between features in atomic-thin films and those in the bulk form. 

In this work, we first have utilized the ability to control both the size and direction of the momentum transfer vector in momentum-dependent electron energy-loss spectroscopy (EELS) of atomically thin MoS$_2$. We characterize the equivalent optical response to reveal the anisotropic properties of the electronic excitations and their layer thickness dependence. EELS has been long recognized as an alternative nonoptical tool to probe electronic structures of semiconductors and has been used to study the valence electron excitation such as band-gap transition \cite{Gu2007, Arenal2005} and plasmonics \cite{Zhou2012, Wachsmuth2013}. In EELS, the momentum transfer vector ($\vec{q}=\vec{k_i}-\vec{k_f}$) plays the role of the polarization vector in optical absorption \cite{Sun2005}, where $\vec{k_i}$ and $\vec{k_f}$ are the wave-vector of the incident and outgoing electrons respectively. The angle- (or momentum-) resolved EELS is thus particularly suited to probe the anisotropy of the electronic transitions because the directions of the momentum transfer can range from being parallel to being perpendicular to the incident direction around the characteristic scattering angle ($\theta_E=E/2E_0$) \cite{Egerton1996}, where $E$ is the energy loss and $E_0$ is the kinetic energy of incident electrons. Our result, taken with the specimen's $c$ axis parallel with the incident beam, can therefore mimic the optical experiments taken both at normal and glancing angle (see Fig. 1a). It has revealed strong difference between the in-plane and out-of-plane polarized response, and has allowed us to track the changes of underlying electronic structures when analyzed in detail with the reported ARPES results \cite{Jin2013, Zhang2014}. In addition, through spectral analysis we not only provide a direct confirmation of the well-known indirect-to-direct gap transition but also determines directly the monolayer's out-of-plane optical gap ($2.4\pm0.2$ eV) which is significantly different from the well-known in-plane gap ($1.8$ eV) \cite{Mak2010} and important to understand energy gap mismatch in vertically stacked heterostructure. The unexpected three-dimensional-like character of the joint density of states (JDOS) of the indirect transition, even in monolayer, has implication on the electronic structure engineering as well as the charge screening effect in MoS$_2$. Our work also provides a further vital experimental check of theoretical calculation \cite{Molina-Sanchez2013} of optical spectra at higher energies, particularly with regard to the layer thickness-dependence as a consequence of quantum confinement effect.

\section{Methods}
Atomically thin MoS$_2$ was prepared through the standard micromechanical exfoliation process, and transferred onto a copper grid with lacey carbon film for TEM observations. No polymer (such as PMMA) was used during the transfer process, which can substantially decrease the possible contamination. 

Monochromated EELS measurements were conducted in a TEM (FEI Titan $60\--300$) equipped with a Gatan Tridiem 865 spectrometer. This microscope was operated at 60 kV in order to reduce the irradiation damage. The attainable energy resolution is less than $140$ meV in the absence of specimens, and this value changes to 0.2 eV under experimental conditions used for MoS$_2$ monolayers. The convergence semiangle of the incident electron beam was set to be less than $\alpha \sim0.3$ mrad to yield a nearly parallel illumination. EEL spectra were recorded in diffraction mode and a rotation holder was used to choose the specific orientation, similar to that used by Wachsmuth \textit{et al.} \cite{Wachsmuth2013}. A selected-area aperture with a diameter of 10 $\mu$m was used for electron diffraction that corresponds to an illumination area with a diameter of 200 nm on the specimen. The scattering geometry of AREELS was limited by a spectrometer entrance aperture (SEA) with a diameter of 2.5 mm (corresponding to 0.54 $\AA^{-1}$). Each $q$-$E$ diagram was recorded for 3 min. All spectra were collected after zero-loss peak alignment and no detectable energy drift ($<0.14$ eV) was observed during EELS acquisition.

\begin{figure}%
\includegraphics[width=1.0\columnwidth]{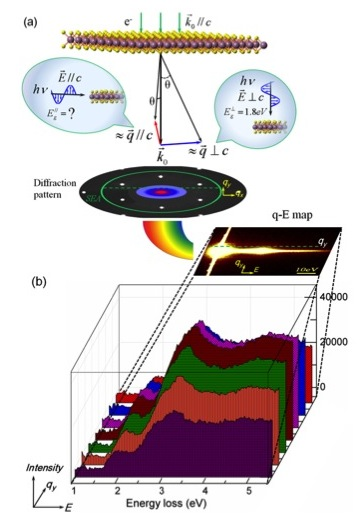}%
\caption{(Colour online) Experimental set-up and the resulting momentum-dependent spectra. (a) Inelastic electron scattering geometry in analogy with polarized optical measuring. The samples in the TEM are atomically thin Mo$S_2$ layers. EELS spectra with in-plane polarization ($\vec{q}$$\perp$$\vec{c}$) dominate the larger-scattering-angle region (blue) and resemble the well-known optical excitation at normal incidence (right inset) onto atomic layers. While the out-of-plane polarized ($\vec{q}$$\parallel$$\vec{c}$) spectra reside in small-scattering-angle region (red) in the momentum space, similar to the unexplored grazing-incidence case of optical wave (left inset). (b) Corresponding momentum dependent spectra extracted from $q$-$E$ diagram with $q_y$ along the $\Gamma \text{M}$ direction in (a).}
\label{FIG1}%
\end{figure}

The valence EEL spectra were acquired from freestanding areas (about 200 nm in size) of high-quality MoS$_2$ monolayers to avoid substrate effect. Figure \ref{FIG1}(a) shows the inelastic scattering geometry and the [0001] zone-axis diffraction pattern of a free-standing monolayer MoS$_2$ at the normal incidence of the electron beam, i.e. parallel with the specimen's $c$ axis. The scattering kinetics dictates that the in-plane polarization ($\vec{q}\perp\vec{c}$) component of the inelastic spectra dominates the large-scattering-angle region (blue). This mode resembles the well-known optical excitation at normal incidence  \cite{Mak2010, Splendiani2010} [right inset in Fig.\ref{FIG1}(a)] due to the similarity of the momentum transfer vector and the electric field vector in the transition matrix element for EELS and optical absorption, respectively. The out-of-plane polarized ($\vec{q}\parallel \vec{c}$)) component resides in the small-scattering-angle region (red) in the momentum space, similar to the unexplored grazing-incidence case of optical spectroscopy (left inset). The two regions are separated by the characteristic angle of $\theta_E$=$E$/$(2E_0) \sim 0.016$ mrad (or $q_E \sim 3.3 \times10^{-4}\AA^{-1}$). 

Superimposed on the diffraction pattern in Fig. \ref{FIG1} is a (green) ring marking the size of the round SEA used in the measurement. The electron energy-loss spectrometer was operated in the energy-dispersive diffraction mode to produce a $q$-$E$ map, with the partially angle integrated energy-loss spectrum in the spectrometer's energy dispersing direction (defined as the $q_x$ direction) while the angular information in the perpendicular direction (defined as the $q_y$ direction) is preserved. The momentum dependent energy-loss map (the $q$-$E$ diagram) shown in Fig.\ref{FIG1}(a) is obtained by summation of 200 individual 1.0 second drift-corrected measurements to enhance the signal. The $q_y$ direction in Fig. \ref{FIG1}(a) is aligned along the $\Gamma \text{M}$ direction using a rotation holder.  The momentum-dependent spectra are line plots as a function of $q_y$ as shown in Fig.\ref{FIG1}(b). Each line plot is an intensity integration along the $q_x$ direction (the energy dispersing direction) over the momentum transfer range limited by the angular size of SEA. For an incident electron beam travelling down the $c$ axis of an uniaxial crystal, the experimental intensity can be written as \cite{Sun2005}:

 \begin{equation}
 I(q_y)=\int_{-\sqrt{q_0^2-q_y^2}}^{\sqrt{q_0^2-q_y^2}}\frac{\text{Im}(\epsilon_{\parallel})q_E^2+\text{Im}(\epsilon_{\perp})(q_x^2+q_y^2)}{\mid \epsilon_{\parallel}q_E^2+\epsilon_{\perp}(q_x^2+q_y^2)\mid^2 } dq_x \bigtriangleup q_y
 \label{EQ1}
 \end{equation}
where $q_0$(=$0.54\AA^{-1}$) is the size of the SEA in the momentum transfer space, $\bigtriangleup q_y$ is the corresponding pixel size in the unit of the scattering momentum space in the $q_y$ direction, $\epsilon$ is the complex dielectric function, and the subscripts denote the polarization directions with respect to the surface normal of the sample. Here atomically thin thickness excludes the possible influence from Cerenkov loss \cite{Festenberg1969}. Even in this partially momentum-integrated form, the anisotropy information should still be visible as we demonstrate below.

For the angular and energy ranges we are interested in, we did not notice any significant difference in the spectrum acquired with $q_y$ aligned along the $\Gamma \text{K}$ direction to that along the $\Gamma \text{M}$ direction (Fig. 3). Wachsmuth \textit{et al.}  \cite{Wachsmuth2013} has used a similar $q$-$E$ mapping approach for graphene, but a specially adopted narrow slit has been used to make the integration over $q_x$ shown in Eq. (\ref{EQ1}) unnecessary to the first approximation. We notice that in-plane anisotropy in their case were only observed at high scattering angle of 72 mrad (near the Brillouin Zone boundary in graphite at $1.2\AA^{-1}$; Ref.[\onlinecite{Wachsmuth2013}]), comparable to the angular range of SEA in our case (54 mrad). Thus we believe that in the low-energy-loss region of interest we can treat in-plane anisotropy to be negligible in our case.

To study the in-plane and out-of-plane anisotropy, we need to resolve the momentum transfer variation in the order of the characteristic angle $\theta_E$ (0.016 mrad for energy loss at 2 eV for 60 keV high-energy electrons). As our beam convergence angle is limited to about 0.3 mrad, due to the need to study finite-sized crystals and the presence of any residual beam divergence, one cannot use the traditional EELS method of a linear slit aperture to provide momentum resolution in the $q_x$ direction. Instead, we have to work with spectra containing mixed contributions and use the well-known numerical angle method \cite{Sun2005, Gu1999} to separate the response from the two orientations, as shown later.
\section{Results}
\begin{figure}%
\includegraphics[width=0.9\columnwidth]{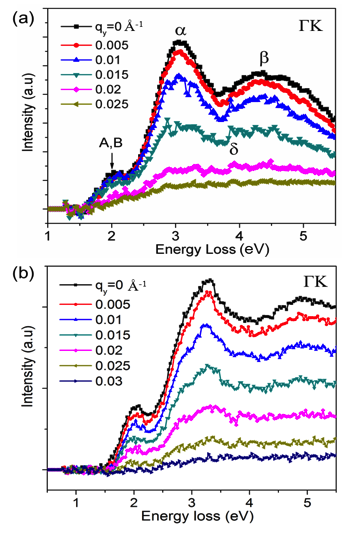}%
\caption{(Colour online) (a) Angle-resolved low loss spectra of monolayer MoS$_2$ with $q_y$ along the $\Gamma K$ direction, in the unit of $\AA^{-1}$. The weak peak at 2.0 eV is due to direct transition of A, B excitons, and the strong peak at 3.1 eV ($\alpha$) and 4.5 eV ($\beta$) result from strong inter-band transitions (van Hove critical points). (b) Spectra of multilayer MoS$_2$ with $q_y$ along $\Gamma K$ direction. The $\delta$ absorption peak arises as $q_y$ increases.}
\label{FIG2}%
\end{figure}

\subsection{Angle-resolved spectra in atomically thin MoS$_2$}
\begin{figure}%
\includegraphics[width=0.85\columnwidth]{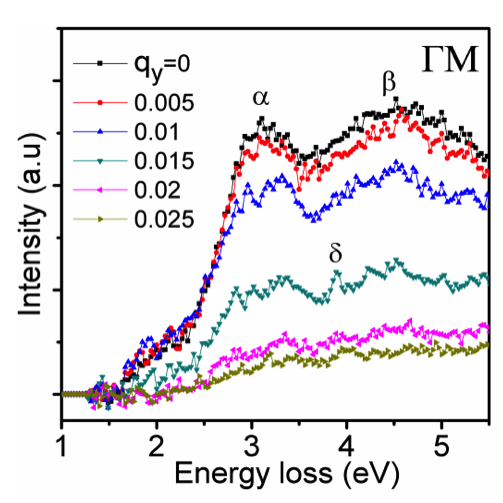}%
\caption{(Colour online) Momentum dependent spectra of monolayer MoS2 with $q_y$ along $\Gamma \text{M}$  direction. The $\alpha$ peak still splits and additional $\delta$ absorption peak appears as $q_y$ increases}%
\label{FIG3}%
\end{figure}

Figure \ref{FIG2}(a) shows the angle-resolved EEL spectra of monolayer MoS$_2$ extracted from the $q$-$E$ map with $q_y$ along the $\Gamma K$ direction. Spectra of multilayer MoS$_2$ with $q_y$ along $\Gamma K$ direction behaves in a similar way to that shown in Fig.\ref{FIG2}(b). Only excitations up to 5 eV are displayed because of our focus on the electronic structure near the absorption edge region. Plasmon excitation at higher energy losses will be discussed in a separate paper. Three noticeable features are observed: a weak transition peaked at 2.0 eV (marked by A, B following the convention of optical spectroscopy \cite{Mak2010} which identifies it as spin-orbit split excitons in the KK' direct transition), a group of strong transition centred at 3.1 eV (the $\alpha$ peak), and a broad peak at 4.5 eV (the $\beta$ peak). A similar result for EELS in monolayer along the $\Gamma M$ direction is also shown in Fig. \ref{FIG3}, indicating that it is a good approximation to treat the in-plane anisotropy as negligible as we have discussed and reasonable to treat dielectric response as being uniaxial as described by Eq. (\ref{EQ1}).

\subsection{Angle-integrated spectra of MoS$_2$ layers with different thicknesses}
\begin{figure}%
\includegraphics[width=0.85\columnwidth]{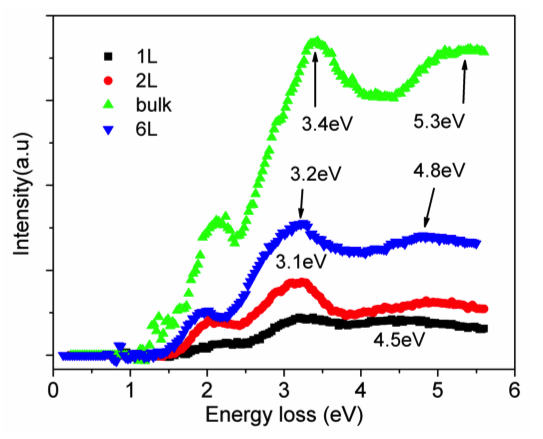}%
\caption{(Colour online) The fine structures of angle-integrated low loss spectra revealing inter-band transitions in MoS$_2$ with different thicknesses.}%
\label{FIG4}%
\end{figure}

In Fig. \ref{FIG4}, we have collected thickness-dependent spectra from ultrathin MoS$_2$ to illustrate the joint density of states and quantum confinement effect of electronic transitions as discussed later. Figure  \ref{FIG4} shows the fine structures of low-loss spectra from MoS$_2$ with variable thicknesses. They are obtained by integrating the ($q_y$, $E$) map over all $q_y$ to improve the statistics for further spectral analysis. 

\section{Discussion}

\begin{figure}%
\includegraphics[width=0.85\columnwidth]{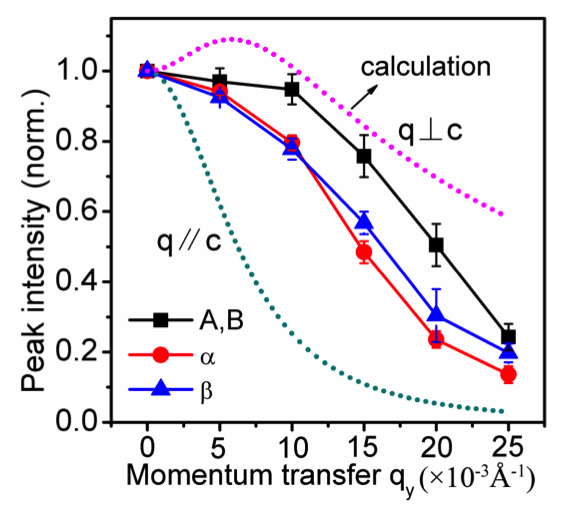}%
\caption{(Colour online) The relative change in peak intensities of A,B, $\alpha$, $\beta$ (at 2 eV, 3.1 eV and 4.5 eV respectively) with momentum transfer $q_y$. Peak intensity at $q_y$ is normalised by its counterpart at $q_y$=0. The experimental data lies between the pure in-plane mode and the pure out-of-plane mode predicted by calculation, illustrating that each spectrum should be a linear superposition of these two components.}%
\label{FIG5}%
\end{figure}
\subsection{Anisotropy in atomically thin MoS$_2$}
The momentum-dependent EEL spectra show strong evidence for the in-plane/out-of-plane anisotropy of the electronic transition in the energy range studied. In Fig. \ref{FIG5}, we have plotted the intensity variations of the three dominant absorptions (A,B; $\alpha$; $\beta$) with the intensity at $q_y$, normalized to their values at $q_y$=0. We expect the out-of-plane component ($\vec{q} \parallel \vec{c}$) to dominate the spectrum at $q_y$=0 and the in-plane ($\vec{q} \perp \vec{c}$) component to dominate at large $q_y$. Although the integration over $q_x$ complicates the spectra interpretation as it mixes the in-plane contribution into the $q_y$=0 spectrum, however, the mixing effect is expected to be negligible as $q_y \gg q_c$. For example, the momentum transfer vector ($\vec{q}$), for the A,B peak centred at 2 eV seen in the $q_y=0.01\AA^{-1}$ spectrum shown in Fig. \ref{FIG2}(a), is determined from electron scattering kinetics to be oriented at least as large as $88^{\circ}$ from the $c$ axis of the sample. The corresponding angle for the 4 eV loss is also higher than $86^{\circ}$, so that the energy-loss spectrum at  $q_y=0.01\AA^{-1}$ over the whole energy range can be interpreted as an in-plane polarized contribution. In general, the mixing of the two polarized contributions is always present but they are expected to show different $q_y$-dependence. For the loss feature centred at 2 eV, the normalized intensity is initially independent of $q_y$, in agreement with the expected dependence of the in-plane contribution. On the other hand, the intensities of the broad $\alpha$, $\beta$ peaks centred at 3.1 eV and 4.5 eV drop more rapidly, consistent with the contribution from the out-of-plane excitation which occurs at small momentum transfer.
\subsection{Separation of in-plane and out-of-plane spectral components}
Another sign for the strong orientation effect of the dielectric response is the disappearance or appearance of certain characteristic spectral features, such as the $\delta$ peak seen at 3.9 eV, as $q_y$ increases [Fig.\ref{FIG2}(a)].  We will make the simple assumption that for the small energy range considered in Fig. \ref{FIG5}, the integrated spectra are linear superposition of the in-plane and out-of-plane components. This is because the integration in Eq. (\ref{EQ1}) can be seen as a linear superposition of the following two components:
\begin{equation}
I_{\perp}=\text{Im}\left( \epsilon_{\perp}\right)\int_{-\sqrt{q_0^2-q_y^2}}^{\sqrt{q_0^2-q_y^2}}\frac{q_x^2+q_y^2}{\mid a(q_x^2+q_y^2)+q_E^2\mid^2}dq_x \bigtriangleup q_y
 \label{EQ2}
\end{equation}
and
\begin{equation}
I_{\parallel}=\text{Im}\left(\epsilon_{\parallel}\right)\int_{-\sqrt{q_0^2-q_y^2}}^{\sqrt{q_0^2-q_y^2}} \frac{q_E^2}{\mid a(q_x^2+q_y^2)+q_E^2\mid^2}dq_x \bigtriangleup q_y
 \label{EQ3}
\end{equation}
where $a=\epsilon_{\perp}/\epsilon_{\parallel}$ is the relative ratio of in-plane to out-of-plane components of the dielectric function.  These two 'pure' components in Fig.5 (dotted line) can be estimated by calculating Eqs. (\ref{EQ2}) and (\ref{EQ3}), where  $a$ is taken as a constant (Fig. \ref{FIG10}) to simplify the calculation. 

Any two experimental spectra $I_1$ and $I_2$ at different $q_y$ can be hybridization of  $I_{\perp}$ and $I_{\parallel}$ components:

\begin{equation}
\begin{align}
I_1=c_1 I_{\perp} + c_2 I_{\parallel} \nonumber, \\
I_2=c_3 I_{\perp} + c_4 I_{\parallel}.
\end{align}
\label{EQ4}
\end{equation}

\begin{figure}%
\includegraphics[width=1.0\columnwidth]{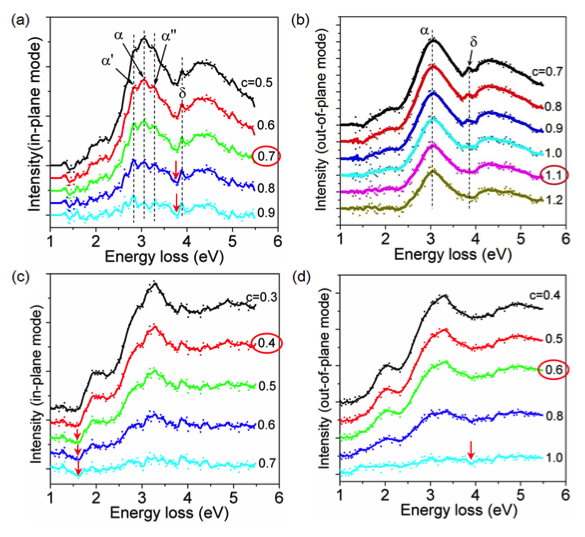}%
\caption{(Colour online) (a,b) The possible in-plane  and out-of-plane  spectral components in monolayer MoS$_2$, based on the subtraction of two momentum-resolved spectra using suitable subtraction factor c. The optimal subtraction factor marked by red circles (c=0.7 for the in-plane component and 1.1 for the out-of-plane component) is based on the presence or absence of the physically realistic $\delta$ feature unique to the in-plane component. It is found that $\alpha$ peak are split into $\alpha'$, $\alpha''$ in the in-plane component while it is a single well-behaved peak in the out-of-plane component. Note that inappropriate subtraction will result in unphysically sharp spectral variation (pit)\cite{Gu1999} in the difference spectrum, as shown by the arrows. (c, d) The in-plane and out-of-plane spectra in multilayer MoS$_2$, extracted from the original EELS data based on the same subtraction procedure.}%
\label{FIG6}%
\end{figure}

Through simple subtraction between the 'hybridized' experimental spectra (Fig. \ref{FIG2} for monolayer and multilayer), we can solve this matrix equation [Eq. \ref{EQ4}] to obtain pure components $I_{\perp}$ and $I_{\parallel}$, by multiplying a constant  $c_1/c_3$ or $c_2/c_4$ to the subtracted spectrum $I_2$. This is similar to the method used by Gu \cite{Gu1999} to separate energy-loss near-edge fine structure (ELNES) at the boundary from that in the bulk by a subtraction. Different trial coefficients of c have been multiplied to the subtracted spectrum to yield a series of possible differential spectra as shown in Fig. \ref{FIG6}.

The key to avoiding a subjective scaling factor determination is to monitoring the abrupt $\delta$ feature which seems to be predominantly in the in-plane directions which are detected in the large '$q_y$' spectra (see Figs. \ref{FIG2} and \ref{FIG3}). The sharp $\delta$ peak, absent in the small angle, is the sole property of the in-plane contribution. For such sharp spectral feature, inappropriate subtraction will result in unphysically sharp spectral variation (pit)\cite{Gu1999} in the difference spectrum, as shown by the arrows in Fig. \ref{FIG6}. As the abrupt $\delta$ feature should be easily recognizable, we use its absence to determine the mostly likely out-of-plane component, marked with a (red) circle in Fig. \ref{FIG6}(b).

Based on this insight, we have extracted pure in-plane or out-of-plane transitions in monolayer and multilayer. The appropriateness of the subtraction can be checked by observing the appearance of fine structure of the $\alpha$ peak in the polarization-resolved spectra. In the polarization-resolved spectra, we see that the in-plane spectrum has a double peak ($\alpha'$, $\alpha''$) structure on the shoulder of the $\alpha$ peak, while the out-of-plane feature has a simple single-peak structure. This provides a simple explanation of the complex $\alpha'$, $\alpha$ and $\alpha''$ peaks in the experimental momentum-dependent spectra as shown in Figs. \ref{FIG2} and \ref{FIG3}. The same method is also used in multilayer system (Fig. \ref{FIG6}(c) and (d)).

\subsection{Polarization-resolved spectroscopy}
\begin{figure}%
\includegraphics[width=1.0\columnwidth]{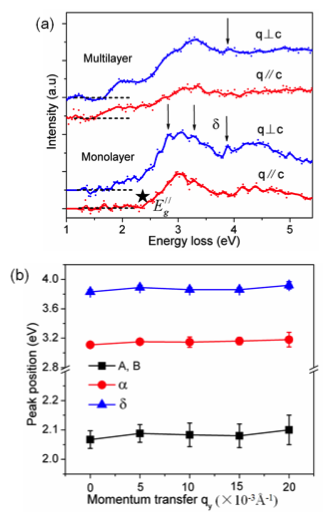}%
\caption{(Colour online)  Polarization dependence of near-gap EELS in atomically-thin MoS$_2$. (a) Absorption spectra of monolayer and multilayer MoS$_2$ at in-plane ($\vec{q} \perp \vec{c}$) and out-of-plane ($\vec{q} \parallel \vec{c}$) polarization. The black star ($\bigstar$) indicates the threshold energy ($2.4\pm0.2 $eV) in out-of-plane polarization which is inaccessible by optical measurements. (b) Dispersion effect of the peak energy (peak position) of major absorption peaks in Fig.\ref{FIG2}a. All lines are drawn as visual aid to the trend seen in the data.}%
\label{FIG7}%
\end{figure}

Fig. \ref{FIG7}(a) shows the resulting in-plane and out-of-plane spectral contributions, using the above 'unbiased' numerical angle method. Here the out-of-plane ($\vec{q} \parallel \vec{c}$) components are similar to the electronic transition probed by grazing-incident light with $\vec{E} \parallel \vec{c}$ polarisation, which is quite difficult to explore on atomic layers by optical means because of the transverse nature of the electromagnetic wave.

The polarization-resolved spectra in monolayer MoS$_2$ immediately suggest that the optical gap in the out-of-plane direction ($E_g^{\parallel}$) has a rather higher value of $2.4\pm0.2$ eV. This means that the direct-gap transition in monolayer seen around 2 eV has an in-plane polarized character. On the other hand, the in-plane and the out-of-plane transitions in the multilayer system near the gap region, which occurs at the smaller energy of $1.5\pm0.2$ eV, seem to have much less obvious polarization dependence, suggesting that the electronic transitions involved in the multilayers have a more three-dimensional(3D)-like character.

The peak energies of the three dominant transitions in the energy range of interest, detected in the monolayer MoS$_2$, display negligible dispersion as shown in Fig.\ref{FIG7}(b). As a result, we can ignore the dispersion effect in our discussions of momentum-dependent spectra. The dispersionless character of the A, B peaks is consistent with the excitonic nature of this spin-orbit split direct transition \cite{Mak2010}. Theoretically the valence band spin-orbit splitting leads to a 0.15 eV difference between the spin-orbital split states of the excitons \cite{Kadantsev2012}. However, the broadening due to phonon scattering and the limited energy resolution of our instrument (0.14 eV) means that we are unable to discriminate these two closely spaced A, B peaks. In fact, the other two sharp transition features we have identified all show negligible variation over the momentum range being probed.

\begin{figure}%
\includegraphics[width=0.95\columnwidth]{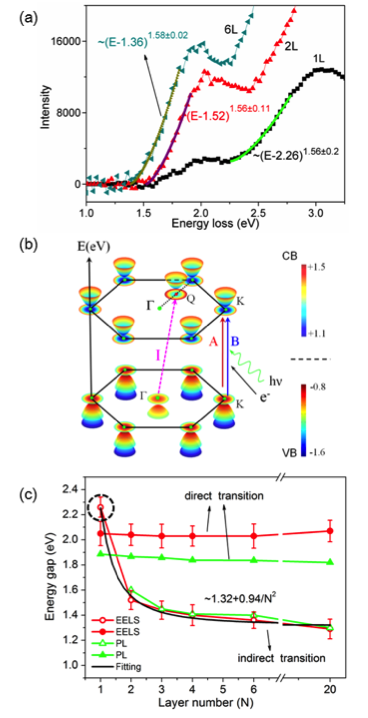}%
\caption{(Colour online)  Dimensionality analysis and quantum confinement effect of electronic transitions. (a) A closer look at band edge transition. Nonlinear fittings of near gap fine structures give the indirect transition energy ($\Gamma \rightarrow \text{Q}$ transition). (b) Schematic illustration of band structure of monolayer MoS$_2$ with global conduction band minimum (CBM) and valence band maximum (VBM) both at K point. The second VBM is located at $\Gamma$-point and the second CBM at Q-point (almost midpoint of straight line $\Gamma \text{K}$). For the sake of brevity, only one Q point paraboloid is drawn (altogether six). The false color discs are the projection of parabolic dispersion to describe the positions of local VBM or CBM. (c) A summarized electronic transition energy between this work, and other reports \protect\cite{Mak2010, Splendiani2010, Kumar2012, Neville1976}. Note the direct transition energies of different layers are extracted from peak energy of A, B exciton transitions. The black dashed circle highlights the indirect transition energy in monolayer system which cannot be easily measured by optical pumping. Nonlinear fitting of indirect transition energies quantitatively confirms the quantum confinement effect.}%
\label{FIG8}%
\end{figure}

The anisotropic property we have observed may be understood in terms of the electronic orbitals of MoS$_2$ (Ref. \cite{Mak2010, Boker2001} which is typical of transition metal dichalcogenides, with the 4d orbitals of Mo situated within the larger energy $\sigma$-$\sigma^*$ gap of bonding and antibonding s-p orbitals \cite{Mattheiss1973, Chhowalla2013}. Because of the trigonal prismatic nature of the S ligand atom arrangement, the 4d orbitals are further split into bonding $e_g$-like upper band involving $d_{xz}$, $d_{yz}$ orbitals and a $t_{2g}$-like lower band involving $d_{z^2}$, $d_{xy}$ and $d_{x^2-y^2}$  orbitals. The hybridization among the symmetry-allowed combination of $d$-orbitals produces the resulting electronic states, but at high symmetry points, it is useful to discuss the states and the electronic transitions in terms of the atomic orbitals involved.  \textit{Ab initio} calculations \cite{Molina-Sanchez2013, Kadantsev2012, Kumar2012} have identified the direct bandgap as the transition at the $K$-point of the first Brillouin Zone (BZ) which has a predominant Mo $d_{x^2-y^2}$ character, and the indirect energy gap arises from the transition from $\Gamma_v$ (local valence band maximum at $\Gamma$-point) to $Q_c$ (local conduction band valley $Q$-point shown in Fig. \ref{FIG8}(b)). The electronic orbital character of $\Gamma_v$ has been identified with Mo-$d_{z^2}$-S-$p_z$ hybrid\cite{Mattheiss1973}. The dipole-allowed transition detected by EELS requires a parity change, thus the direct $d$-$d$ transition at the $K$ point should strictly only be allowed for  $\vec{q} \perp \vec{c}$   polarization as we have observed in the monolayer case [Fig.7(a)] and the indirect transition from $\Gamma_v$ should only be allowed in $\vec{q} \parallel \vec{c}$ polarization. Our result for the multilayer suggests that the second selection rule is relaxed.

The non-bonding nature of the $d_{x^2-y^2}$ orbital involved in the threshold transition also means that the monolayer MoS$_2$ could be a more durable ultrathin photodetector \cite{Yin2012, Lee2012} as it is less likely to suffer from photo-bleaching effect. On the other hand, this polarization-specific optical response offers a way to improve the performance of MoS$_2$-based optoelectronic\cite{Zhang2012, Radisavljevic2011, Radisavljevic2011a, Baugher2013, Yin2012, Lee2012} devices in photovoltaics. For example, the out-of-plane dipole transitions in monolayer provide a distinctive possibility to more efficiently utilize the photons above the in-plane-polarized gap 1.8 eV and enhance the photoconversion of solar energy engineered through either adjusting interlayer coupling or surface adsorption.

\subsection{Higher energy excitation}
The wide energy range covered by the EELS method allows us to easily see the higher energy excitation at 3.1-3.5 eV. This is predicted by the accurate theoretical quasiparticle calculation at the level of GW approximation for the energy level and Bethe-Salpeter equation for the absorption spectrum \cite{Molina-Sanchez2013} and has not been observed by the optical method so far. Our analysis also reveals interesting anisotropy. For example, the sharp in-plane $\delta$ feature can be associated with the direct in-plane K-K' excitation from the spin-orbital split valence band maximum to the upper conduction band minimum about the K point. The quasiparticle-excitation calculation of 1H-MoS$_2$ and few-layer 2H-MoS$_2$ indicate that the $\alpha$ peak at 3.1 eV (Fig. \ref{EQ2}a) should arise from the transition between the parallel conduction band and valence band around $Q$ point, where high joint density of states is involved \cite{Molina-Sanchez2013}. Interestingly, the $\alpha$ (3 eV) peak structure changes from a single peak in the out-of-plane direction to the split twin peak structure in the in-plane direction. These subtle anisotropy changes are worth further investigation. Nevertheless, the absorption at 3 eV should also be highly efficient because it is not restricted by the anisotropy-imposed selection rules.

\subsection{Dimensionality analysis of joint density-of-states and layer-thickness dependence}
Spectral analysis, based on fitting the power-law ($E-E_g$)$^n$ to the angle integrated spectra, has traditionally been used to determine the character of the near-threshold low-energy absorption. Examples have shown that diamond has an indirect bandgap\cite{Rafferty1998} and GaAs has a direct bandgap \cite{Batson1986}. In arriving this, Brown and Rafferty have shown that electronic excitation intensity is described as \cite{Rafferty1998}:

\begin{equation}
I=\frac{d^2\sigma}{dE d\Omega}\propto \left\{ 
\begin{aligned} 
\text{JDOS}(E)\\
(E-E_g)\text{JDOS}(E) 
\end{aligned}
\right. ,
\label{EQ6}
\end{equation}
where JDOS($E$) is the joint density of states (JDOS). The upper equation in (\ref{EQ6}) applies for direct transitions, and lower one for indirect transitions. For the 3D semiconductors, the JDOS for a parabolic dispersion follows $(E-Eg)^{0.5}$. Hence in the 3D case, the intensity for direct transition presents the power law with exponent of $n=0.5$ and indirect transition $n=1.5$, verified experimentally for a number of 3D semiconducting materials. We have extended their argument on the energy-dependence of the spectral intensity for direct and indirect transitions in two dimensional electronic systems where JDOS has an exponent $n=0$. Hence the corresponding power-law exponents of the spectral analysis should, in principle, be $n=0$ (direct) and $n=1$ (indirect) respectively.

We investigate the dimensionality of the joint-density of states (JDOS) involved through power-law analysis of the spectral variations. The fitted exponents for the threshold transitions in bilayers and multilayers [shown in Fig. \ref{FIG8}(a)] are all close to 1.5. Together with the polarization-insensitive energy gap transition shown in Fig.\ref{FIG7}(a), they demonstrate that the transition involved has a three-dimensional-like character. The observation of three-dimensional-like dispersion in the few layer systems (2L, 3L, 6L) is consistent with the strong dispersion of the valence band (dominated by S-$p_z$ states) in the $\Gamma A$ orientation \cite{Boker2001, Coehoorn1987} and has previously been understood as interlayer coupling of the S-$p_z$ orbitals across the van der Waal spacing. This 3D-like character is also in accord with the electrostatic screening experiment \cite{Castellanos-Gomez2013}.

We then follow the evolution of the indirect transition as a function of the layer thickness, even after the cross-over from indirect-to-direct gap transition [Fig. \ref{FIG8}(a)]. This is because the overlapping direct transition in the monolayer system should have a power-law dependence of ($(E-E_g)^0$), similar to the JDOS of the two-dimensional character \cite{Liang1973} of the band structure near the K-point. Although near the threshold, the spectrum does not follow such power-law due to the strong excitonic peak \cite{Molina-Sanchez2013, Neville1976}, we expect that such a power-law dependence prevails above the bandgap region where the excitonic effect is absent. Indeed, such a flat absorption band has been well known experimentally in the related TMDs \cite{Liang1973}. We therefore identify the rising absorption above direct transition at 2 eV in monolayer with the same transition responsible for the indirect $\Gamma \rightarrow Q$ transition (Fig. \ref{FIG8}(b)) in multilayered MoS$_2$. This identification is supported by recent photoconductivity measurements indicating that the transition in this energy range is associated with excitation into the conduction band at the $Q$ point \cite{Kozawa2014}. Giving support to this identification, the power-fitting returns a spectral intensity exponent of 1.56$\pm$0.20. The deduced transition energy is 2.26$\pm$0.06 eV, corresponding to a 3D-like indirect transition in monolayer, as the case in the multilayer system. In comparison, the $\Gamma \rightarrow Q$ indirect transition is predicted to be $2.0\pm0.1$ eV from LDA calculation \cite{Molina-Sanchez2013, Kumar2012} and 2.5$\pm$ 0.2eV from the more accurate quasiparticle calculation \cite{Molina-Sanchez2013} in monolayer system. 

    The observation of 3D-like JDOS associated with the indirect transition in the monolayer case cannot be explained as interlayer coupling across the van-der-Waals spacing as it is absent. We suggest that it is because the confinement potential is 'soft' in the sense that it rises slowly with the distance from the surfaces. It is well known that if the confinement potential along the out-of-plane direction is abrupt ('hard-edge') as found in semiconductor quantum well structures, then we expect the allowed $k_z$ values of the carrier wavefunctions to take discreet values. If however, the confinement potential approaches the vacuum level more gradually ('soft-edge') as one moves away from the monolayer, then it is possible that the allowed $k_z$ can take quasi-continuous values and JDOS may follow $(E-E_g)^{0.5}$ as it approaches the vacuum level, as in the case of Rydberg atoms. 
    
    Our 3D-like JDOS result is consistent with the assumption made by Castellanos-Gomez \textit{et al.} \cite{Castellanos-Gomez2013} in explaining their electrostatic screening data in monolayer MoS$_2$.  In their study of the electrostatic screening by means of electrostatic force microscopy in combination with a non-linear Thomas-Fermi theory, they find that a continuum model of decoupled layers, which satisfactorily reproduces the electrostatic screening for graphene and graphite, cannot account for the experimental observations in MoS$_2$. A three-dimensional model with an interlayer hopping parameter can, on the other hand, successfully account for the observed electric field screening by MoS$_2$ nanolayers, pointing out the important role of the interlayer coupling in the screening of MoS$_2$.
    
\subsection{Thickness-dependent transition energies}

The aforementioned three-dimensional nature of the indirect transition provides a logical explanation for the observed strong layer dependence of the threshold energy for the indirect transition. Figure \ref{FIG8}(c) summarizes the transition threshold energies in MoS$_2$ with different thicknesses which are generated from the previous quantitative power-law spectral analysis. Table \ref{TAB1} compares our EELS-derived transition gap values, the reported photoluminescence and calculation results, with all showing consistent trend with layer thickness. It must be pointed out that the indirect gap of multilayer systems evolves into the direct gap of the monolayer MoS$_2$, when the number of layers is reduced. However, EELS allows us to track the indirect transition even as its threshold energy is now shifted beyond the direct gap transition energy. To test independently our assignment of the indirect electronic transitions in the monolayer, we have found that the threshold energy of the indirect transition as a function of the layer number, N, can be fitted by the expression $E_g=E_B+A/N^2$ as shown by the fitting curve in Fig. \ref{FIG8}(c). $E_B$ is $1.32\pm0.01$ eV and corresponds to the indirect gap of bulk MoS$_2$. The quantitatively well-fitted black line is consistent with the size dependence of the energies in a classic one-dimensional potential well and further verifies the physical picture of quantum confinement effect in layered MoS$_2$.

\begin{table*}[h!]
  \centering
  \caption{A comparison of energy gaps for both indirect transitions ($\Gamma \rightarrow$Q) and direct transitions (K$\rightarrow$K') using EELS, PL or DFT methods.}
  \label{TAB1}
  \begin{ruledtabular}
  \begin{tabular}{ccccc}
      & Indirect energy gap (eV) & & Direct energy gap (eV)&\\
  \end{tabular}
  \centering
  \begin{tabular}{ccccccc}
    Layers & EELS & PL\cite{Mak2010} & DFT\cite{Kadantsev2012, Kumar2012} & EELS & PL \cite{Mak2010} & DFT\cite{Kadantsev2012, Kumar2012b}\\
    \hline
    1 & 2.26$\pm$0.11 & a & 2.1 & 2.05$\pm$0.10 & 1.89 & 1.79 \\
    2 & 1.52$\pm$0.11 & 1.60 & 1.6 & 2.04$\pm$0.08 & 1.87 & 1.78\\
    3 & 1.44$\pm$0.11 & 1.45 & 1.5 & 2.03$\pm$0.09 & 1.86 & 1.76 \\
    4 & 1.40$\pm$0.11 & 1.41 & 1.3 & 2.03$\pm$0.08 & 1.84 & 1.74 \\
    6 & 1.36$\pm$0.11 & 1.40 & 1.1 & 2.03$\pm$0.09 & 1.84 & 1.73\\
    bulk & 1.29$\pm$0.11 & 1.30 & 0.9 & 2.03$\pm$0.08 & 1.82 & 1.73\\
  \end{tabular}
  \end{ruledtabular}
  \footnote{The energy of the indirect transition in monolayer system is not measurable using PL because it is higher than the energy of the direct gap transition.}
\end{table*}

Calculations by Molina-Sanchez \textit{et al.} and Kumar and Ahluwalia \cite{Molina-Sanchez2013, Kumar2012} predict a downward shift of the $\Gamma_v$-point when the layer thickness is reduced. This has been corroborated by ARPES measurements although the bandwidth of the valence band is smaller in experiment because of the substrate effect \cite{Jin2013}. To date, there is no experimental information about the movement of the conduction band structure. Therefore, we use our determination of the energies for the direct and indirect transitions (in Fig. \ref{FIG8}(c)) to map out the movement of the conduction band valley Q$_c$ (Fig. \ref{FIG8}(b)). The ARPES data suggests that the $\Gamma_v$-point was downward shifted from its bulk value to that for the monolayer by 0.7 eV. From our data, the evolution of the indirect transition energy suggests that the local Q$_c$-point moves upward by a much smaller amount (0.3 eV) comparable with the theoretical calculation \cite{Molina-Sanchez2013}. This may arise from different atomic orbital hybridizations:  the former $\Gamma_v$ (mainly Mo-$d_{z^2}$-S-$p_z$) is more sensitive to interlayer coupling than the Q$_c$ ($d_{z^2}$, $d_{xy}$ and $d_{x^2-y^2}$ orbitals). This asymmetry has consequence for many physical properties, such as changes in the effective mass of electrons and holes which are important for the transport in few-layer MoS$_2$ systems.

\subsection{Effect of radiation damage on EELS fine structures}

\begin{figure}%
\includegraphics[width=0.8\columnwidth]{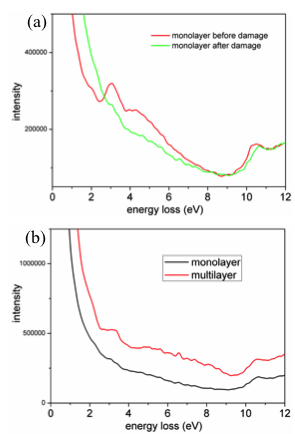}%
\caption{(Colour online) ). (a) Low-loss EEL spectra from a fresh MoS$_2$  monolayer (in red) and that after exposure to electron beam irradiation for 4 hours at an acceleration voltage of 60 kV. (b) Low-loss EEL spectra recorded from monolayer and multilayer MoS$_2$ after long-time electron beam irradiation.}%
\label{FIG9}%
\end{figure}

By comparing the EEL spectra recorded from the monolayer MoS$_2$ before and after the beam damage as shown in Fig.\ref{FIG9}, we found that the damage associated with long-term electron beam irradiation could lead to the disappearance of A, B excitons peaked at 2 eV and the strong inter-band transitions ($\alpha$ and $\beta$ at 3.1 eV and 4.5 eV respectively) in monolayer MoS$_2$. For multilayer specimens, the beam damage after extended exposure (about 4 hours here) will obviously weaken the fine structures in their EEL spectra. Results from such a comparison here demonstrate that the low-loss fine structures in EEL spectra are mainly coming from the transitions between electronic bands of intact crystals. 

\section{Summary}
In conclusion, we showed that electronic excitation in MoS$_2$ presents not only the well-known indirect- to direct-gap transition, but also less-anisotropic to highly anisotropic response, as the thickness is reduced to monolayer. The latter favors the optical absorption by normally incident light, hence may be partially responsible for the enhanced photoabsorption seen in monolayer MoS$_2$. In particular, the well-known 1.8 eV direct gap in monolayer MoS$_2$ is shown to be entirely in-plane polarized, consistent with the 2D character of the $d_{x^2-y^2}$-like orbitals involved, while the out-of-plane-polarized energy gap is much larger at $2.4\pm0.2$ eV. Such an extremely anisotropic optical and electronic property in the monolayer system is important for the development of efficient photovoltaic and photocatalysis applications. It may give rise to novel interlayer transport property in layer-stacking heterostructures.

    We also showed that the joint-density-of-states of the indirect $\Gamma$-Q transition is 3D-like in character, even in the monolayer case. We propose that this is due to the 'soft' edge nature of the confinement potential, which makes the electronic states involved easily tunable by vertical stacking. We showed experimentally that the threshold energies of the indirect $\Gamma$-Q transition follows the well-known quantum confinement scaling relationship, in contrast with the layer-thickness independence of the K-K' direct transition. Our investigation presents a systematic and comprehensive insight into the physics of this new semiconductor and may also benefit orientation-dependent applications in optoelectronics and heterostructured electronics.

\begin{figure}[t!]%
\includegraphics[width=0.95\columnwidth]{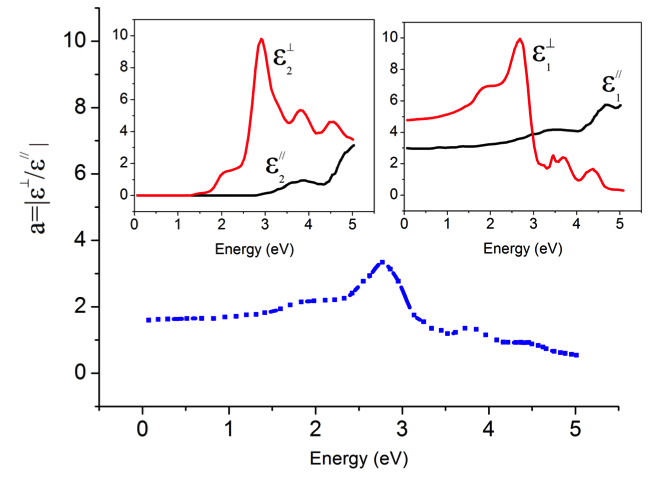}%
\caption{(Colour online). Dielectric functions and their ratio ($a=\mid$$\epsilon^{\perp}/\epsilon^{\parallel}$$\mid$), the inset dielectric constants extracted from Ref. \cite{Kumar2012}. It shows small variation of "$a$" as a function of energy.}%
\label{FIG10}%
\end{figure}

\section*{Acknowledgements}
This work is financially supported by the National Basic Research Program of China (Grants No. 2014CB932500 and No. 2015CB921000) and National Science Foundation of China (Grants No. 51222202 and No. 51472215). The research reported in this paper was partially supported by King Abdullah University of Science and Technology (KAUST). J.Y. acknowledges Pao Yu-Kong International Foundation for a Visiting Chair Professorship in ZJU and EPSRC and Royal Society for partial support. Prof. Ray Egerton and Dr. He Tian are kindly acknowledged for critical reading of the paper.  

JH and KL contribute equally to this work.
\section*{Appendix}
The rational behind our assumption that each spectrum can be linear superposition of these two components stems from the observation that the factor '$a$' only appears in the denominator in Eq. \ref{EQ3} as the coefficient of the two quadratic terms, so the effect of small variation as a function of energy over the narrow energy range of interest can be negligible as no strong plasmon.  Using the theoretical result, the spectral variation of $a$ can be estimated and the result is shown in Fig. \ref{FIG10} resonance is involved.

\bibliography{MoS2}

\begin{thebibliography}{41}%
\makeatletter
\providecommand \@ifxundefined [1]{%
 \@ifx{#1\undefined}
}%
\providecommand \@ifnum [1]{%
 \ifnum #1\expandafter \@firstoftwo
 \else \expandafter \@secondoftwo
 \fi
}%
\providecommand \@ifx [1]{%
 \ifx #1\expandafter \@firstoftwo
 \else \expandafter \@secondoftwo
 \fi
}%
\providecommand \natexlab [1]{#1}%
\providecommand \enquote  [1]{``#1''}%
\providecommand \bibnamefont  [1]{#1}%
\providecommand \bibfnamefont [1]{#1}%
\providecommand \citenamefont [1]{#1}%
\providecommand \href@noop [0]{\@secondoftwo}%
\providecommand \href [0]{\begingroup \@sanitize@url \@href}%
\providecommand \@href[1]{\@@startlink{#1}\@@href}%
\providecommand \@@href[1]{\endgroup#1\@@endlink}%
\providecommand \@sanitize@url [0]{\catcode `\\12\catcode `\$12\catcode
  `\&12\catcode `\#12\catcode `\^12\catcode `\_12\catcode `\%12\relax}%
\providecommand \@@startlink[1]{}%
\providecommand \@@endlink[0]{}%
\providecommand \url  [0]{\begingroup\@sanitize@url \@url }%
\providecommand \@url [1]{\endgroup\@href {#1}{\urlprefix }}%
\providecommand \urlprefix  [0]{URL }%
\providecommand \Eprint [0]{\href }%
\providecommand \doibase [0]{http://dx.doi.org/}%
\providecommand \selectlanguage [0]{\@gobble}%
\providecommand \bibinfo  [0]{\@secondoftwo}%
\providecommand \bibfield  [0]{\@secondoftwo}%
\providecommand \translation [1]{[#1]}%
\providecommand \BibitemOpen [0]{}%
\providecommand \bibitemStop [0]{}%
\providecommand \bibitemNoStop [0]{.\EOS\space}%
\providecommand \EOS [0]{\spacefactor3000\relax}%
\providecommand \BibitemShut  [1]{\csname bibitem#1\endcsname}%
\let\auto@bib@innerbib\@empty
\bibitem [{\citenamefont {Wang}\ \emph {et~al.}(2012)\citenamefont {Wang},
  \citenamefont {Kalantar-Zadeh}, \citenamefont {Kis}, \citenamefont
  {Coleman},\ and\ \citenamefont {Strano}}]{Wang2012}%
  \BibitemOpen
  \bibfield  {author} {\bibinfo {author} {\bibfnamefont {Q.~H.}\ \bibnamefont
  {Wang}}, \bibinfo {author} {\bibfnamefont {K.}~\bibnamefont
  {Kalantar-Zadeh}}, \bibinfo {author} {\bibfnamefont {A.}~\bibnamefont {Kis}},
  \bibinfo {author} {\bibfnamefont {J.~N.}\ \bibnamefont {Coleman}}, \ and\
  \bibinfo {author} {\bibfnamefont {M.~S.}\ \bibnamefont {Strano}},\ }\href
  {\doibase 10.1038/nnano.2012.193} {\bibfield  {journal} {\bibinfo  {journal}
  {Nature nanotechnology}\ }\textbf {\bibinfo {volume} {7}},\ \bibinfo {pages}
  {699} (\bibinfo {year} {2012})}\BibitemShut {NoStop}%
\bibitem [{\citenamefont {Ataca}\ \emph {et~al.}(2012)\citenamefont {Ataca},
  \citenamefont {Şahin},\ and\ \citenamefont {Ciraci}}]{Ataca2012}%
  \BibitemOpen
  \bibfield  {author} {\bibinfo {author} {\bibfnamefont {C.}~\bibnamefont
  {Ataca}}, \bibinfo {author} {\bibfnamefont {H.}~\bibnamefont {Şahin}}, \
  and\ \bibinfo {author} {\bibfnamefont {S.}~\bibnamefont {Ciraci}},\ }\href
  {\doibase 10.1021/jp212558p} {\bibfield  {journal} {\bibinfo  {journal} {The
  Journal of Physical Chemistry C}\ }\textbf {\bibinfo {volume} {116}},\
  \bibinfo {pages} {8983} (\bibinfo {year} {2012})}\BibitemShut {NoStop}%
\bibitem [{\citenamefont {Zhang}\ \emph {et~al.}(2012)\citenamefont {Zhang},
  \citenamefont {Ye}, \citenamefont {Matsuhashi},\ and\ \citenamefont
  {Iwasa}}]{Zhang2012}%
  \BibitemOpen
  \bibfield  {author} {\bibinfo {author} {\bibfnamefont {Y.}~\bibnamefont
  {Zhang}}, \bibinfo {author} {\bibfnamefont {J.}~\bibnamefont {Ye}}, \bibinfo
  {author} {\bibfnamefont {Y.}~\bibnamefont {Matsuhashi}}, \ and\ \bibinfo
  {author} {\bibfnamefont {Y.}~\bibnamefont {Iwasa}},\ }\href {\doibase
  10.1021/nl2021575} {\bibfield  {journal} {\bibinfo  {journal} {Nano letters}\
  }\textbf {\bibinfo {volume} {12}},\ \bibinfo {pages} {1136} (\bibinfo {year}
  {2012})}\BibitemShut {NoStop}%
\bibitem [{\citenamefont {Radisavljevic}\ \emph
  {et~al.}(2011{\natexlab{a}})\citenamefont {Radisavljevic}, \citenamefont
  {Whitwick},\ and\ \citenamefont {Kis}}]{Radisavljevic2011}%
  \BibitemOpen
  \bibfield  {author} {\bibinfo {author} {\bibfnamefont {B.}~\bibnamefont
  {Radisavljevic}}, \bibinfo {author} {\bibfnamefont {M.~B.}\ \bibnamefont
  {Whitwick}}, \ and\ \bibinfo {author} {\bibfnamefont {A.}~\bibnamefont
  {Kis}},\ }\href {\doibase 10.1021/nn203715c} {\bibfield  {journal} {\bibinfo
  {journal} {ACS nano}\ }\textbf {\bibinfo {volume} {5}},\ \bibinfo {pages}
  {9934} (\bibinfo {year} {2011}{\natexlab{a}})}\BibitemShut {NoStop}%
\bibitem [{\citenamefont {Radisavljevic}\ \emph
  {et~al.}(2011{\natexlab{b}})\citenamefont {Radisavljevic}, \citenamefont
  {Radenovic}, \citenamefont {Brivio}, \citenamefont {Giacometti},\ and\
  \citenamefont {Kis}}]{Radisavljevic2011a}%
  \BibitemOpen
  \bibfield  {author} {\bibinfo {author} {\bibfnamefont {B.}~\bibnamefont
  {Radisavljevic}}, \bibinfo {author} {\bibfnamefont {A.}~\bibnamefont
  {Radenovic}}, \bibinfo {author} {\bibfnamefont {J.}~\bibnamefont {Brivio}},
  \bibinfo {author} {\bibfnamefont {V.}~\bibnamefont {Giacometti}}, \ and\
  \bibinfo {author} {\bibfnamefont {A.}~\bibnamefont {Kis}},\ }\href {\doibase
  10.1038/nnano.2010.279} {\bibfield  {journal} {\bibinfo  {journal} {Nature
  nanotechnology}\ }\textbf {\bibinfo {volume} {6}},\ \bibinfo {pages} {147}
  (\bibinfo {year} {2011}{\natexlab{b}})}\BibitemShut {NoStop}%
\bibitem [{\citenamefont {Baugher}\ \emph {et~al.}(2013)\citenamefont
  {Baugher}, \citenamefont {Churchill}, \citenamefont {Yang},\ and\
  \citenamefont {Jarillo-Herrero}}]{Baugher2013}%
  \BibitemOpen
  \bibfield  {author} {\bibinfo {author} {\bibfnamefont {B.~W.~H.}\
  \bibnamefont {Baugher}}, \bibinfo {author} {\bibfnamefont {H.~O.~H.}\
  \bibnamefont {Churchill}}, \bibinfo {author} {\bibfnamefont {Y.}~\bibnamefont
  {Yang}}, \ and\ \bibinfo {author} {\bibfnamefont {P.}~\bibnamefont
  {Jarillo-Herrero}},\ }\href {\doibase 10.1021/nl401916s} {\bibfield
  {journal} {\bibinfo  {journal} {Nano letters}\ }\textbf {\bibinfo {volume}
  {13}},\ \bibinfo {pages} {4212} (\bibinfo {year} {2013})}\BibitemShut
  {NoStop}%
\bibitem [{\citenamefont {Yin}\ \emph {et~al.}(2012)\citenamefont {Yin},
  \citenamefont {Li}, \citenamefont {Li}, \citenamefont {Jiang}, \citenamefont
  {Shi}, \citenamefont {Sun}, \citenamefont {Lu}, \citenamefont {Zhang},
  \citenamefont {Chen},\ and\ \citenamefont {Zhang}}]{Yin2012}%
  \BibitemOpen
  \bibfield  {author} {\bibinfo {author} {\bibfnamefont {Z.}~\bibnamefont
  {Yin}}, \bibinfo {author} {\bibfnamefont {H.}~\bibnamefont {Li}}, \bibinfo
  {author} {\bibfnamefont {H.}~\bibnamefont {Li}}, \bibinfo {author}
  {\bibfnamefont {L.}~\bibnamefont {Jiang}}, \bibinfo {author} {\bibfnamefont
  {Y.}~\bibnamefont {Shi}}, \bibinfo {author} {\bibfnamefont {Y.}~\bibnamefont
  {Sun}}, \bibinfo {author} {\bibfnamefont {G.}~\bibnamefont {Lu}}, \bibinfo
  {author} {\bibfnamefont {Q.}~\bibnamefont {Zhang}}, \bibinfo {author}
  {\bibfnamefont {X.}~\bibnamefont {Chen}}, \ and\ \bibinfo {author}
  {\bibfnamefont {H.}~\bibnamefont {Zhang}},\ }\href {\doibase
  10.1021/nn2024557} {\bibfield  {journal} {\bibinfo  {journal} {ACS nano}\
  }\textbf {\bibinfo {volume} {6}},\ \bibinfo {pages} {74} (\bibinfo {year}
  {2012})}\BibitemShut {NoStop}%
\bibitem [{\citenamefont {Lee}\ \emph {et~al.}(2012)\citenamefont {Lee},
  \citenamefont {Min}, \citenamefont {Chang}, \citenamefont {Park},
  \citenamefont {Nam}, \citenamefont {Kim}, \citenamefont {Kim}, \citenamefont
  {Ryu},\ and\ \citenamefont {Im}}]{Lee2012}%
  \BibitemOpen
  \bibfield  {author} {\bibinfo {author} {\bibfnamefont {H.~S.}\ \bibnamefont
  {Lee}}, \bibinfo {author} {\bibfnamefont {S.-W.}\ \bibnamefont {Min}},
  \bibinfo {author} {\bibfnamefont {Y.-G.}\ \bibnamefont {Chang}}, \bibinfo
  {author} {\bibfnamefont {M.~K.}\ \bibnamefont {Park}}, \bibinfo {author}
  {\bibfnamefont {T.}~\bibnamefont {Nam}}, \bibinfo {author} {\bibfnamefont
  {H.}~\bibnamefont {Kim}}, \bibinfo {author} {\bibfnamefont {J.~H.}\
  \bibnamefont {Kim}}, \bibinfo {author} {\bibfnamefont {S.}~\bibnamefont
  {Ryu}}, \ and\ \bibinfo {author} {\bibfnamefont {S.}~\bibnamefont {Im}},\
  }\href {\doibase 10.1021/nl301485q} {\bibfield  {journal} {\bibinfo
  {journal} {Nano letters}\ }\textbf {\bibinfo {volume} {12}},\ \bibinfo
  {pages} {3695} (\bibinfo {year} {2012})}\BibitemShut {NoStop}%
\bibitem [{\citenamefont {Zeng}\ \emph {et~al.}(2012)\citenamefont {Zeng},
  \citenamefont {Dai}, \citenamefont {Yao}, \citenamefont {Xiao},\ and\
  \citenamefont {Cui}}]{Zeng2012}%
  \BibitemOpen
  \bibfield  {author} {\bibinfo {author} {\bibfnamefont {H.}~\bibnamefont
  {Zeng}}, \bibinfo {author} {\bibfnamefont {J.}~\bibnamefont {Dai}}, \bibinfo
  {author} {\bibfnamefont {W.}~\bibnamefont {Yao}}, \bibinfo {author}
  {\bibfnamefont {D.}~\bibnamefont {Xiao}}, \ and\ \bibinfo {author}
  {\bibfnamefont {X.}~\bibnamefont {Cui}},\ }\href {\doibase
  10.1038/nnano.2012.95} {\bibfield  {journal} {\bibinfo  {journal} {Nature
  nanotechnology}\ }\textbf {\bibinfo {volume} {7}},\ \bibinfo {pages} {490}
  (\bibinfo {year} {2012})}\BibitemShut {NoStop}%
\bibitem [{\citenamefont {Mak}\ \emph {et~al.}(2012)\citenamefont {Mak},
  \citenamefont {He}, \citenamefont {Shan},\ and\ \citenamefont
  {Heinz}}]{Mak2012}%
  \BibitemOpen
  \bibfield  {author} {\bibinfo {author} {\bibfnamefont {K.~F.}\ \bibnamefont
  {Mak}}, \bibinfo {author} {\bibfnamefont {K.}~\bibnamefont {He}}, \bibinfo
  {author} {\bibfnamefont {J.}~\bibnamefont {Shan}}, \ and\ \bibinfo {author}
  {\bibfnamefont {T.~F.}\ \bibnamefont {Heinz}},\ }\href {\doibase
  10.1038/nnano.2012.96} {\bibfield  {journal} {\bibinfo  {journal} {Nature
  nanotechnology}\ }\textbf {\bibinfo {volume} {7}},\ \bibinfo {pages} {494}
  (\bibinfo {year} {2012})}\BibitemShut {NoStop}%
\bibitem [{\citenamefont {Li}\ and\ \citenamefont {Galli}(2007)}]{Li2007a}%
  \BibitemOpen
  \bibfield  {author} {\bibinfo {author} {\bibfnamefont {T.}~\bibnamefont
  {Li}}\ and\ \bibinfo {author} {\bibfnamefont {G.}~\bibnamefont {Galli}},\
  }\href {\doibase 10.1021/jp075424v} {\bibfield  {journal} {\bibinfo
  {journal} {Journal of Physical Chemistry C}\ }\textbf {\bibinfo {volume}
  {111}},\ \bibinfo {pages} {16192} (\bibinfo {year} {2007})}\BibitemShut
  {NoStop}%
\bibitem [{\citenamefont {Leb{\`{e}}gue}\ and\ \citenamefont
  {Eriksson}(2009)}]{Lebegue2009}%
  \BibitemOpen
  \bibfield  {author} {\bibinfo {author} {\bibfnamefont {S.}~\bibnamefont
  {Leb{\`{e}}gue}}\ and\ \bibinfo {author} {\bibfnamefont {O.}~\bibnamefont
  {Eriksson}},\ }\href {\doibase 10.1103/PhysRevB.79.115409} {\bibfield
  {journal} {\bibinfo  {journal} {Physical Review B}\ }\textbf {\bibinfo
  {volume} {79}},\ \bibinfo {pages} {115409} (\bibinfo {year}
  {2009})}\BibitemShut {NoStop}%
\bibitem [{\citenamefont {Mak}\ \emph {et~al.}(2010)\citenamefont {Mak},
  \citenamefont {Lee}, \citenamefont {Hone}, \citenamefont {Shan},\ and\
  \citenamefont {Heinz}}]{Mak2010}%
  \BibitemOpen
  \bibfield  {author} {\bibinfo {author} {\bibfnamefont {K.~F.}\ \bibnamefont
  {Mak}}, \bibinfo {author} {\bibfnamefont {C.}~\bibnamefont {Lee}}, \bibinfo
  {author} {\bibfnamefont {J.}~\bibnamefont {Hone}}, \bibinfo {author}
  {\bibfnamefont {J.}~\bibnamefont {Shan}}, \ and\ \bibinfo {author}
  {\bibfnamefont {T.~F.}\ \bibnamefont {Heinz}},\ }\href {\doibase
  10.1103/PhysRevLett.105.136805} {\bibfield  {journal} {\bibinfo  {journal}
  {Physical Review Letters}\ }\textbf {\bibinfo {volume} {105}},\ \bibinfo
  {pages} {136805} (\bibinfo {year} {2010})}\BibitemShut {NoStop}%
\bibitem [{\citenamefont {Splendiani}\ \emph {et~al.}(2010)\citenamefont
  {Splendiani}, \citenamefont {Sun}, \citenamefont {Zhang}, \citenamefont {Li},
  \citenamefont {Kim}, \citenamefont {Chim}, \citenamefont {Galli},\ and\
  \citenamefont {Wang}}]{Splendiani2010}%
  \BibitemOpen
  \bibfield  {author} {\bibinfo {author} {\bibfnamefont {A.}~\bibnamefont
  {Splendiani}}, \bibinfo {author} {\bibfnamefont {L.}~\bibnamefont {Sun}},
  \bibinfo {author} {\bibfnamefont {Y.}~\bibnamefont {Zhang}}, \bibinfo
  {author} {\bibfnamefont {T.}~\bibnamefont {Li}}, \bibinfo {author}
  {\bibfnamefont {J.}~\bibnamefont {Kim}}, \bibinfo {author} {\bibfnamefont
  {C.-Y.}\ \bibnamefont {Chim}}, \bibinfo {author} {\bibfnamefont
  {G.}~\bibnamefont {Galli}}, \ and\ \bibinfo {author} {\bibfnamefont
  {F.}~\bibnamefont {Wang}},\ }\href {\doibase 10.1021/nl903868w} {\bibfield
  {journal} {\bibinfo  {journal} {Nano letters}\ }\textbf {\bibinfo {volume}
  {10}},\ \bibinfo {pages} {1271} (\bibinfo {year} {2010})}\BibitemShut
  {NoStop}%
\bibitem [{\citenamefont {Jin}\ \emph {et~al.}(2013)\citenamefont {Jin},
  \citenamefont {Yeh}, \citenamefont {Zaki}, \citenamefont {Zhang},
  \citenamefont {Sadowski}, \citenamefont {Al-Mahboob}, \citenamefont {van~der
  Zande}, \citenamefont {Chenet}, \citenamefont {Dadap}, \citenamefont
  {Herman}, \citenamefont {Sutter}, \citenamefont {Hone},\ and\ \citenamefont
  {Osgood}}]{Jin2013}%
  \BibitemOpen
  \bibfield  {author} {\bibinfo {author} {\bibfnamefont {W.}~\bibnamefont
  {Jin}}, \bibinfo {author} {\bibfnamefont {P.-c.}\ \bibnamefont {Yeh}},
  \bibinfo {author} {\bibfnamefont {N.}~\bibnamefont {Zaki}}, \bibinfo {author}
  {\bibfnamefont {D.}~\bibnamefont {Zhang}}, \bibinfo {author} {\bibfnamefont
  {J.~T.}\ \bibnamefont {Sadowski}}, \bibinfo {author} {\bibfnamefont
  {A.}~\bibnamefont {Al-Mahboob}}, \bibinfo {author} {\bibfnamefont {A.~M.}\
  \bibnamefont {van~der Zande}}, \bibinfo {author} {\bibfnamefont {D.~a.}\
  \bibnamefont {Chenet}}, \bibinfo {author} {\bibfnamefont {J.~I.}\
  \bibnamefont {Dadap}}, \bibinfo {author} {\bibfnamefont {I.~P.}\ \bibnamefont
  {Herman}}, \bibinfo {author} {\bibfnamefont {P.}~\bibnamefont {Sutter}},
  \bibinfo {author} {\bibfnamefont {J.}~\bibnamefont {Hone}}, \ and\ \bibinfo
  {author} {\bibfnamefont {R.~M.}\ \bibnamefont {Osgood}},\ }\href {\doibase
  10.1103/PhysRevLett.111.106801} {\bibfield  {journal} {\bibinfo  {journal}
  {Physical Review Letters}\ }\textbf {\bibinfo {volume} {111}},\ \bibinfo
  {pages} {106801} (\bibinfo {year} {2013})}\BibitemShut {NoStop}%
\bibitem [{\citenamefont {Zhang}\ \emph {et~al.}(2014)\citenamefont {Zhang},
  \citenamefont {Chang}, \citenamefont {Zhou}, \citenamefont {Cui},
  \citenamefont {Yan}, \citenamefont {Liu}, \citenamefont {Schmitt},
  \citenamefont {Lee}, \citenamefont {Moore}, \citenamefont {Chen},
  \citenamefont {Lin}, \citenamefont {Jeng}, \citenamefont {Mo}, \citenamefont
  {Hussain}, \citenamefont {Bansil},\ and\ \citenamefont {Shen}}]{Zhang2014}%
  \BibitemOpen
  \bibfield  {author} {\bibinfo {author} {\bibfnamefont {Y.}~\bibnamefont
  {Zhang}}, \bibinfo {author} {\bibfnamefont {T.-r.}\ \bibnamefont {Chang}},
  \bibinfo {author} {\bibfnamefont {B.}~\bibnamefont {Zhou}}, \bibinfo {author}
  {\bibfnamefont {Y.-t.}\ \bibnamefont {Cui}}, \bibinfo {author} {\bibfnamefont
  {H.}~\bibnamefont {Yan}}, \bibinfo {author} {\bibfnamefont {Z.}~\bibnamefont
  {Liu}}, \bibinfo {author} {\bibfnamefont {F.}~\bibnamefont {Schmitt}},
  \bibinfo {author} {\bibfnamefont {J.}~\bibnamefont {Lee}}, \bibinfo {author}
  {\bibfnamefont {R.}~\bibnamefont {Moore}}, \bibinfo {author} {\bibfnamefont
  {Y.}~\bibnamefont {Chen}}, \bibinfo {author} {\bibfnamefont {H.}~\bibnamefont
  {Lin}}, \bibinfo {author} {\bibfnamefont {H.-t.}\ \bibnamefont {Jeng}},
  \bibinfo {author} {\bibfnamefont {S.-k.}\ \bibnamefont {Mo}}, \bibinfo
  {author} {\bibfnamefont {Z.}~\bibnamefont {Hussain}}, \bibinfo {author}
  {\bibfnamefont {A.}~\bibnamefont {Bansil}}, \ and\ \bibinfo {author}
  {\bibfnamefont {Z.-x.}\ \bibnamefont {Shen}},\ }\href {\doibase
  10.1038/nnano.2013.277} {\bibfield  {journal} {\bibinfo  {journal} {Nature
  nanotechnology}\ }\textbf {\bibinfo {volume} {9}},\ \bibinfo {pages} {111}
  (\bibinfo {year} {2014})}\BibitemShut {NoStop}%
\bibitem [{\citenamefont {Geim}\ and\ \citenamefont
  {Grigorieva}(2013)}]{Geim2013}%
  \BibitemOpen
  \bibfield  {author} {\bibinfo {author} {\bibfnamefont {A.~K.}\ \bibnamefont
  {Geim}}\ and\ \bibinfo {author} {\bibfnamefont {I.~V.}\ \bibnamefont
  {Grigorieva}},\ }\href {\doibase 10.1038/nature12385} {\bibfield  {journal}
  {\bibinfo  {journal} {Nature}\ }\textbf {\bibinfo {volume} {499}},\ \bibinfo
  {pages} {419} (\bibinfo {year} {2013})},\ \Eprint
  {http://arxiv.org/abs/1307.6718} {arXiv:1307.6718} \BibitemShut {NoStop}%
\bibitem [{\citenamefont {Georgiou}\ \emph {et~al.}(2013)\citenamefont
  {Georgiou}, \citenamefont {Jalil}, \citenamefont {Belle}, \citenamefont
  {Britnell}, \citenamefont {Gorbachev}, \citenamefont {Morozov}, \citenamefont
  {Kim}, \citenamefont {Gholinia}, \citenamefont {Haigh}, \citenamefont
  {Makarovsky}, \citenamefont {Eaves}, \citenamefont {Ponomarenko},
  \citenamefont {Geim}, \citenamefont {Novoselov},\ and\ \citenamefont
  {Mishchenko}}]{Georgiou2013}%
  \BibitemOpen
  \bibfield  {author} {\bibinfo {author} {\bibfnamefont {T.}~\bibnamefont
  {Georgiou}}, \bibinfo {author} {\bibfnamefont {R.}~\bibnamefont {Jalil}},
  \bibinfo {author} {\bibfnamefont {B.~D.}\ \bibnamefont {Belle}}, \bibinfo
  {author} {\bibfnamefont {L.}~\bibnamefont {Britnell}}, \bibinfo {author}
  {\bibfnamefont {R.~V.}\ \bibnamefont {Gorbachev}}, \bibinfo {author}
  {\bibfnamefont {S.~V.}\ \bibnamefont {Morozov}}, \bibinfo {author}
  {\bibfnamefont {Y.-J.}\ \bibnamefont {Kim}}, \bibinfo {author} {\bibfnamefont
  {A.}~\bibnamefont {Gholinia}}, \bibinfo {author} {\bibfnamefont {S.~J.}\
  \bibnamefont {Haigh}}, \bibinfo {author} {\bibfnamefont {O.}~\bibnamefont
  {Makarovsky}}, \bibinfo {author} {\bibfnamefont {L.}~\bibnamefont {Eaves}},
  \bibinfo {author} {\bibfnamefont {L.~a.}\ \bibnamefont {Ponomarenko}},
  \bibinfo {author} {\bibfnamefont {A.~K.}\ \bibnamefont {Geim}}, \bibinfo
  {author} {\bibfnamefont {K.~S.}\ \bibnamefont {Novoselov}}, \ and\ \bibinfo
  {author} {\bibfnamefont {A.}~\bibnamefont {Mishchenko}},\ }\href {\doibase
  10.1038/nnano.2012.224} {\bibfield  {journal} {\bibinfo  {journal} {Nature
  nanotechnology}\ }\textbf {\bibinfo {volume} {8}},\ \bibinfo {pages} {100}
  (\bibinfo {year} {2013})}\BibitemShut {NoStop}%
\bibitem [{\citenamefont {Britnell}\ \emph {et~al.}(2013)\citenamefont
  {Britnell}, \citenamefont {Ribeiro}, \citenamefont {Eckmann}, \citenamefont
  {Jalil}, \citenamefont {Belle}, \citenamefont {Mishchenko}, \citenamefont
  {Kim}, \citenamefont {Gorbachev}, \citenamefont {Georgiou}, \citenamefont
  {Morozov}, \citenamefont {Grigorenko}, \citenamefont {Geim}, \citenamefont
  {Casiraghi}, \citenamefont {{Castro Neto}},\ and\ \citenamefont
  {Novoselov}}]{Britnell2013}%
  \BibitemOpen
  \bibfield  {author} {\bibinfo {author} {\bibfnamefont {L.}~\bibnamefont
  {Britnell}}, \bibinfo {author} {\bibfnamefont {R.~M.}\ \bibnamefont
  {Ribeiro}}, \bibinfo {author} {\bibfnamefont {A.}~\bibnamefont {Eckmann}},
  \bibinfo {author} {\bibfnamefont {R.}~\bibnamefont {Jalil}}, \bibinfo
  {author} {\bibfnamefont {B.~D.}\ \bibnamefont {Belle}}, \bibinfo {author}
  {\bibfnamefont {A.}~\bibnamefont {Mishchenko}}, \bibinfo {author}
  {\bibfnamefont {Y.-J.}\ \bibnamefont {Kim}}, \bibinfo {author} {\bibfnamefont
  {R.~V.}\ \bibnamefont {Gorbachev}}, \bibinfo {author} {\bibfnamefont
  {T.}~\bibnamefont {Georgiou}}, \bibinfo {author} {\bibfnamefont {S.~V.}\
  \bibnamefont {Morozov}}, \bibinfo {author} {\bibfnamefont {A.~N.}\
  \bibnamefont {Grigorenko}}, \bibinfo {author} {\bibfnamefont {A.~K.}\
  \bibnamefont {Geim}}, \bibinfo {author} {\bibfnamefont {C.}~\bibnamefont
  {Casiraghi}}, \bibinfo {author} {\bibfnamefont {A.~H.}\ \bibnamefont {{Castro
  Neto}}}, \ and\ \bibinfo {author} {\bibfnamefont {K.~S.}\ \bibnamefont
  {Novoselov}},\ }\href {\doibase 10.1126/science.1235547} {\bibfield
  {journal} {\bibinfo  {journal} {Science (New York, N.Y.)}\ }\textbf {\bibinfo
  {volume} {340}},\ \bibinfo {pages} {1311} (\bibinfo {year}
  {2013})}\BibitemShut {NoStop}%
\bibitem [{\citenamefont {Liang}(1973)}]{Liang1973}%
  \BibitemOpen
  \bibfield  {author} {\bibinfo {author} {\bibfnamefont {W.}~\bibnamefont
  {Liang}},\ }\href {http://iopscience.iop.org/0022-3719/6/3/018} {\bibfield
  {journal} {\bibinfo  {journal} {Journal of Physics C: Solid State Physics}\
  }\textbf {\bibinfo {volume} {6}},\ \bibinfo {pages} {551－565} (\bibinfo
  {year} {1973})}\BibitemShut {NoStop}%
\bibitem [{\citenamefont {Gu}\ \emph {et~al.}(2007)\citenamefont {Gu},
  \citenamefont {Srot}, \citenamefont {Sigle}, \citenamefont {Koch},
  \citenamefont {van Aken}, \citenamefont {Scholz}, \citenamefont {Thapa},
  \citenamefont {Kirchner}, \citenamefont {Jetter},\ and\ \citenamefont
  {R{\"{u}}hle}}]{Gu2007}%
  \BibitemOpen
  \bibfield  {author} {\bibinfo {author} {\bibfnamefont {L.}~\bibnamefont
  {Gu}}, \bibinfo {author} {\bibfnamefont {V.}~\bibnamefont {Srot}}, \bibinfo
  {author} {\bibfnamefont {W.}~\bibnamefont {Sigle}}, \bibinfo {author}
  {\bibfnamefont {C.}~\bibnamefont {Koch}}, \bibinfo {author} {\bibfnamefont
  {P.}~\bibnamefont {van Aken}}, \bibinfo {author} {\bibfnamefont
  {F.}~\bibnamefont {Scholz}}, \bibinfo {author} {\bibfnamefont
  {S.}~\bibnamefont {Thapa}}, \bibinfo {author} {\bibfnamefont
  {C.}~\bibnamefont {Kirchner}}, \bibinfo {author} {\bibfnamefont
  {M.}~\bibnamefont {Jetter}}, \ and\ \bibinfo {author} {\bibfnamefont
  {M.}~\bibnamefont {R{\"{u}}hle}},\ }\href {\doibase
  10.1103/PhysRevB.75.195214} {\bibfield  {journal} {\bibinfo  {journal}
  {Physical Review B}\ }\textbf {\bibinfo {volume} {75}},\ \bibinfo {pages}
  {195214} (\bibinfo {year} {2007})}\BibitemShut {NoStop}%
\bibitem [{\citenamefont {Arenal}\ \emph {et~al.}(2005)\citenamefont {Arenal},
  \citenamefont {St{\'{e}}phan}, \citenamefont {Kociak}, \citenamefont
  {Taverna}, \citenamefont {Loiseau},\ and\ \citenamefont
  {Colliex}}]{Arenal2005}%
  \BibitemOpen
  \bibfield  {author} {\bibinfo {author} {\bibfnamefont {R.}~\bibnamefont
  {Arenal}}, \bibinfo {author} {\bibfnamefont {O.}~\bibnamefont
  {St{\'{e}}phan}}, \bibinfo {author} {\bibfnamefont {M.}~\bibnamefont
  {Kociak}}, \bibinfo {author} {\bibfnamefont {D.}~\bibnamefont {Taverna}},
  \bibinfo {author} {\bibfnamefont {A.}~\bibnamefont {Loiseau}}, \ and\
  \bibinfo {author} {\bibfnamefont {C.}~\bibnamefont {Colliex}},\ }\href
  {\doibase 10.1103/PhysRevLett.95.127601} {\bibfield  {journal} {\bibinfo
  {journal} {Physical Review Letters}\ }\textbf {\bibinfo {volume} {95}},\
  \bibinfo {pages} {127601} (\bibinfo {year} {2005})}\BibitemShut {NoStop}%
\bibitem [{\citenamefont {Zhou}\ \emph {et~al.}(2012)\citenamefont {Zhou},
  \citenamefont {Lee}, \citenamefont {Nanda}, \citenamefont {Pantelides},
  \citenamefont {Pennycook},\ and\ \citenamefont {Idrobo}}]{Zhou2012}%
  \BibitemOpen
  \bibfield  {author} {\bibinfo {author} {\bibfnamefont {W.}~\bibnamefont
  {Zhou}}, \bibinfo {author} {\bibfnamefont {J.}~\bibnamefont {Lee}}, \bibinfo
  {author} {\bibfnamefont {J.}~\bibnamefont {Nanda}}, \bibinfo {author}
  {\bibfnamefont {S.~T.}\ \bibnamefont {Pantelides}}, \bibinfo {author}
  {\bibfnamefont {S.~J.}\ \bibnamefont {Pennycook}}, \ and\ \bibinfo {author}
  {\bibfnamefont {J.-C.}\ \bibnamefont {Idrobo}},\ }\href {\doibase
  10.1038/nnano.2011.252} {\bibfield  {journal} {\bibinfo  {journal} {Nature
  nanotechnology}\ }\textbf {\bibinfo {volume} {7}},\ \bibinfo {pages} {161}
  (\bibinfo {year} {2012})}\BibitemShut {NoStop}%
\bibitem [{\citenamefont {Wachsmuth}\ \emph {et~al.}(2013)\citenamefont
  {Wachsmuth}, \citenamefont {Hambach}, \citenamefont {Kinyanjui},
  \citenamefont {Guzzo}, \citenamefont {Benner},\ and\ \citenamefont
  {Kaiser}}]{Wachsmuth2013}%
  \BibitemOpen
  \bibfield  {author} {\bibinfo {author} {\bibfnamefont {P.}~\bibnamefont
  {Wachsmuth}}, \bibinfo {author} {\bibfnamefont {R.}~\bibnamefont {Hambach}},
  \bibinfo {author} {\bibfnamefont {M.~K.}\ \bibnamefont {Kinyanjui}}, \bibinfo
  {author} {\bibfnamefont {M.}~\bibnamefont {Guzzo}}, \bibinfo {author}
  {\bibfnamefont {G.}~\bibnamefont {Benner}}, \ and\ \bibinfo {author}
  {\bibfnamefont {U.}~\bibnamefont {Kaiser}},\ }\href {\doibase
  10.1103/PhysRevB.88.075433} {\bibfield  {journal} {\bibinfo  {journal}
  {Physical Review B}\ }\textbf {\bibinfo {volume} {88}},\ \bibinfo {pages}
  {075433} (\bibinfo {year} {2013})}\BibitemShut {NoStop}%
\bibitem [{\citenamefont {Sun}\ and\ \citenamefont {Yuan}(2005)}]{Sun2005}%
  \BibitemOpen
  \bibfield  {author} {\bibinfo {author} {\bibfnamefont {Y.}~\bibnamefont
  {Sun}}\ and\ \bibinfo {author} {\bibfnamefont {J.}~\bibnamefont {Yuan}},\
  }\href {\doibase 10.1103/PhysRevB.71.125109} {\bibfield  {journal} {\bibinfo
  {journal} {Physical Review B}\ }\textbf {\bibinfo {volume} {71}},\ \bibinfo
  {pages} {125109} (\bibinfo {year} {2005})}\BibitemShut {NoStop}%
\bibitem [{\citenamefont {Egerton}(1996)}]{Egerton1996}%
  \BibitemOpen
  \bibfield  {author} {\bibinfo {author} {\bibfnamefont {R.~F.}\ \bibnamefont
  {Egerton}},\ }\href@noop {} {\emph {\bibinfo {title} {{Electron energy loss
  spectroscopy in the electron microscope}}}}\ (\bibinfo  {publisher}
  {Plenum},\ \bibinfo {address} {New York},\ \bibinfo {year}
  {1996})\BibitemShut {NoStop}%
\bibitem [{\citenamefont {Molina-S{\'{a}}nchez}\ \emph
  {et~al.}(2013)\citenamefont {Molina-S{\'{a}}nchez}, \citenamefont {Sangalli},
  \citenamefont {Hummer}, \citenamefont {Marini},\ and\ \citenamefont
  {Wirtz}}]{Molina-Sanchez2013}%
  \BibitemOpen
  \bibfield  {author} {\bibinfo {author} {\bibfnamefont {A.}~\bibnamefont
  {Molina-S{\'{a}}nchez}}, \bibinfo {author} {\bibfnamefont {D.}~\bibnamefont
  {Sangalli}}, \bibinfo {author} {\bibfnamefont {K.}~\bibnamefont {Hummer}},
  \bibinfo {author} {\bibfnamefont {A.}~\bibnamefont {Marini}}, \ and\ \bibinfo
  {author} {\bibfnamefont {L.}~\bibnamefont {Wirtz}},\ }\href {\doibase
  10.1103/PhysRevB.88.045412} {\bibfield  {journal} {\bibinfo  {journal}
  {Physical Review B}\ }\textbf {\bibinfo {volume} {88}},\ \bibinfo {pages}
  {045412} (\bibinfo {year} {2013})},\ \Eprint {http://arxiv.org/abs/1306.4257}
  {arXiv:1306.4257} \BibitemShut {NoStop}%
\bibitem [{\citenamefont {Festenberg}(1969)}]{Festenberg1969}%
  \BibitemOpen
  \bibfield  {author} {\bibinfo {author} {\bibfnamefont {C.~V.}\ \bibnamefont
  {Festenberg}},\ }\href {\doibase 10.1007/BF01394892} {\bibfield  {journal}
  {\bibinfo  {journal} {Zeitschrift f{\"{u}}r Physik}\ }\textbf {\bibinfo
  {volume} {227}},\ \bibinfo {pages} {453} (\bibinfo {year}
  {1969})}\BibitemShut {NoStop}%
\bibitem [{\citenamefont {Gu}(1999)}]{Gu1999}%
  \BibitemOpen
  \bibfield  {author} {\bibinfo {author} {\bibfnamefont {H.}~\bibnamefont
  {Gu}},\ }\href {\doibase 10.1016/S0304-3991(98)00083-7} {\bibfield  {journal}
  {\bibinfo  {journal} {Ultramicroscopy}\ }\textbf {\bibinfo {volume} {76}},\
  \bibinfo {pages} {159} (\bibinfo {year} {1999})}\BibitemShut {NoStop}%
\bibitem [{\citenamefont {Kadantsev}\ and\ \citenamefont
  {Hawrylak}(2012)}]{Kadantsev2012}%
  \BibitemOpen
  \bibfield  {author} {\bibinfo {author} {\bibfnamefont {E.~S.}\ \bibnamefont
  {Kadantsev}}\ and\ \bibinfo {author} {\bibfnamefont {P.}~\bibnamefont
  {Hawrylak}},\ }\href {\doibase 10.1016/j.ssc.2012.02.005} {\bibfield
  {journal} {\bibinfo  {journal} {Solid State Communications}\ }\textbf
  {\bibinfo {volume} {152}},\ \bibinfo {pages} {909} (\bibinfo {year}
  {2012})}\BibitemShut {NoStop}%
\bibitem [{\citenamefont {B{\"{o}}ker}\ \emph {et~al.}(2001)\citenamefont
  {B{\"{o}}ker}, \citenamefont {Severin}, \citenamefont {M{\"{u}}ller},
  \citenamefont {Janowitz}, \citenamefont {Manzke}, \citenamefont {Vo{\ss}},
  \citenamefont {Kr{\"{u}}ger}, \citenamefont {Mazur},\ and\ \citenamefont
  {Pollmann}}]{Boker2001}%
  \BibitemOpen
  \bibfield  {author} {\bibinfo {author} {\bibfnamefont {T.}~\bibnamefont
  {B{\"{o}}ker}}, \bibinfo {author} {\bibfnamefont {R.}~\bibnamefont
  {Severin}}, \bibinfo {author} {\bibfnamefont {A.}~\bibnamefont
  {M{\"{u}}ller}}, \bibinfo {author} {\bibfnamefont {C.}~\bibnamefont
  {Janowitz}}, \bibinfo {author} {\bibfnamefont {R.}~\bibnamefont {Manzke}},
  \bibinfo {author} {\bibfnamefont {D.}~\bibnamefont {Vo{\ss}}}, \bibinfo
  {author} {\bibfnamefont {P.}~\bibnamefont {Kr{\"{u}}ger}}, \bibinfo {author}
  {\bibfnamefont {A.}~\bibnamefont {Mazur}}, \ and\ \bibinfo {author}
  {\bibfnamefont {J.}~\bibnamefont {Pollmann}},\ }\href {\doibase
  10.1103/PhysRevB.64.235305} {\bibfield  {journal} {\bibinfo  {journal}
  {Physical Review B}\ }\textbf {\bibinfo {volume} {64}},\ \bibinfo {pages}
  {235305} (\bibinfo {year} {2001})}\BibitemShut {NoStop}%
\bibitem [{\citenamefont {Mattheiss}(1973)}]{Mattheiss1973}%
  \BibitemOpen
  \bibfield  {author} {\bibinfo {author} {\bibfnamefont {L.~F.}\ \bibnamefont
  {Mattheiss}},\ }\href {\doibase 10.1103/PhysRevB.8.3719} {\bibfield
  {journal} {\bibinfo  {journal} {Physical Review B}\ }\textbf {\bibinfo
  {volume} {8}},\ \bibinfo {pages} {3719} (\bibinfo {year} {1973})}\BibitemShut
  {NoStop}%
\bibitem [{\citenamefont {Chhowalla}\ \emph {et~al.}(2013)\citenamefont
  {Chhowalla}, \citenamefont {Shin}, \citenamefont {Eda}, \citenamefont {Li},
  \citenamefont {Loh},\ and\ \citenamefont {Zhang}}]{Chhowalla2013}%
  \BibitemOpen
  \bibfield  {author} {\bibinfo {author} {\bibfnamefont {M.}~\bibnamefont
  {Chhowalla}}, \bibinfo {author} {\bibfnamefont {H.~S.}\ \bibnamefont {Shin}},
  \bibinfo {author} {\bibfnamefont {G.}~\bibnamefont {Eda}}, \bibinfo {author}
  {\bibfnamefont {L.-j.}\ \bibnamefont {Li}}, \bibinfo {author} {\bibfnamefont
  {K.~P.}\ \bibnamefont {Loh}}, \ and\ \bibinfo {author} {\bibfnamefont
  {H.}~\bibnamefont {Zhang}},\ }\href {\doibase 10.1038/nchem.1589} {\bibfield
  {journal} {\bibinfo  {journal} {Nature chemistry}\ }\textbf {\bibinfo
  {volume} {5}},\ \bibinfo {pages} {263} (\bibinfo {year} {2013})}\BibitemShut
  {NoStop}%
\bibitem [{\citenamefont {Kumar}\ and\ \citenamefont
  {Ahluwalia}(2012{\natexlab{a}})}]{Kumar2012}%
  \BibitemOpen
  \bibfield  {author} {\bibinfo {author} {\bibfnamefont {A.}~\bibnamefont
  {Kumar}}\ and\ \bibinfo {author} {\bibfnamefont {P.}~\bibnamefont
  {Ahluwalia}},\ }\href {\doibase 10.1016/j.matchemphys.2012.05.055} {\bibfield
   {journal} {\bibinfo  {journal} {Materials Chemistry and Physics}\ }\textbf
  {\bibinfo {volume} {135}},\ \bibinfo {pages} {755} (\bibinfo {year}
  {2012}{\natexlab{a}})}\BibitemShut {NoStop}%
\bibitem [{\citenamefont {Neville}\ and\ \citenamefont
  {Evans}(1976)}]{Neville1976}%
  \BibitemOpen
  \bibfield  {author} {\bibinfo {author} {\bibfnamefont {R.~A.}\ \bibnamefont
  {Neville}}\ and\ \bibinfo {author} {\bibfnamefont {B.~L.}\ \bibnamefont
  {Evans}},\ }\href {\doibase 10.1002/pssb.2220730227} {\bibfield  {journal}
  {\bibinfo  {journal} {physica status solidi (b)}\ }\textbf {\bibinfo {volume}
  {73}},\ \bibinfo {pages} {597} (\bibinfo {year} {1976})}\BibitemShut
  {NoStop}%
\bibitem [{\citenamefont {Rafferty}\ and\ \citenamefont
  {Brown}(1998)}]{Rafferty1998}%
  \BibitemOpen
  \bibfield  {author} {\bibinfo {author} {\bibfnamefont {B.}~\bibnamefont
  {Rafferty}}\ and\ \bibinfo {author} {\bibfnamefont {L.~M.}\ \bibnamefont
  {Brown}},\ }\href@noop {} {\bibfield  {journal} {\bibinfo  {journal} {Phys.
  Rev. B}\ }\textbf {\bibinfo {volume} {58}},\ \bibinfo {pages} {326} (\bibinfo
  {year} {1998})}\BibitemShut {NoStop}%
\bibitem [{\citenamefont {Batson}\ \emph {et~al.}(1986)\citenamefont {Batson},
  \citenamefont {Kavanagh}, \citenamefont {Woodall},\ and\ \citenamefont
  {Mayer}}]{Batson1986}%
  \BibitemOpen
  \bibfield  {author} {\bibinfo {author} {\bibfnamefont {P.~E.}\ \bibnamefont
  {Batson}}, \bibinfo {author} {\bibfnamefont {K.~L.}\ \bibnamefont
  {Kavanagh}}, \bibinfo {author} {\bibfnamefont {J.~M.~J.}\ \bibnamefont
  {Woodall}}, \ and\ \bibinfo {author} {\bibfnamefont {J.~W.~J.}\ \bibnamefont
  {Mayer}},\ }\href {\doibase 10.1103/PhysRevLett.57.2729} {\bibfield
  {journal} {\bibinfo  {journal} {Physical review letters}\ }\textbf {\bibinfo
  {volume} {57}},\ \bibinfo {pages} {2729} (\bibinfo {year}
  {1986})}\BibitemShut {NoStop}%
\bibitem [{\citenamefont {Coehoorn}\ \emph {et~al.}(1987)\citenamefont
  {Coehoorn}, \citenamefont {Haas},\ and\ \citenamefont
  {de~Groot}}]{Coehoorn1987}%
  \BibitemOpen
  \bibfield  {author} {\bibinfo {author} {\bibfnamefont {R.}~\bibnamefont
  {Coehoorn}}, \bibinfo {author} {\bibfnamefont {C.}~\bibnamefont {Haas}}, \
  and\ \bibinfo {author} {\bibfnamefont {R.~A.}\ \bibnamefont {de~Groot}},\
  }\href {\doibase 10.1103/PhysRevB.35.6203} {\bibfield  {journal} {\bibinfo
  {journal} {Physical Review B}\ }\textbf {\bibinfo {volume} {35}},\ \bibinfo
  {pages} {6203} (\bibinfo {year} {1987})}\BibitemShut {NoStop}%
\bibitem [{\citenamefont {Castellanos-Gomez}\ \emph {et~al.}(2013)\citenamefont
  {Castellanos-Gomez}, \citenamefont {Cappelluti}, \citenamefont
  {Rold{\'{a}}n}, \citenamefont {Agra{\"{\i}}t}, \citenamefont {Guinea},\ and\
  \citenamefont {Rubio-Bollinger}}]{Castellanos-Gomez2013}%
  \BibitemOpen
  \bibfield  {author} {\bibinfo {author} {\bibfnamefont {A.}~\bibnamefont
  {Castellanos-Gomez}}, \bibinfo {author} {\bibfnamefont {E.}~\bibnamefont
  {Cappelluti}}, \bibinfo {author} {\bibfnamefont {R.}~\bibnamefont
  {Rold{\'{a}}n}}, \bibinfo {author} {\bibfnamefont {N.}~\bibnamefont
  {Agra{\"{\i}}t}}, \bibinfo {author} {\bibfnamefont {F.}~\bibnamefont
  {Guinea}}, \ and\ \bibinfo {author} {\bibfnamefont {G.}~\bibnamefont
  {Rubio-Bollinger}},\ }\href {\doibase 10.1002/adma.201203731} {\bibfield
  {journal} {\bibinfo  {journal} {Advanced Materials}\ }\textbf {\bibinfo
  {volume} {25}},\ \bibinfo {pages} {899} (\bibinfo {year} {2013})}\BibitemShut
  {NoStop}%
\bibitem [{\citenamefont {Kozawa}\ \emph {et~al.}(2014)\citenamefont {Kozawa},
  \citenamefont {Kumar}, \citenamefont {Carvalho}, \citenamefont {{Kumar
  Amara}}, \citenamefont {Zhao}, \citenamefont {Wang}, \citenamefont {Toh},
  \citenamefont {Ribeiro}, \citenamefont {{Castro Neto}}, \citenamefont
  {Matsuda},\ and\ \citenamefont {Eda}}]{Kozawa2014}%
  \BibitemOpen
  \bibfield  {author} {\bibinfo {author} {\bibfnamefont {D.}~\bibnamefont
  {Kozawa}}, \bibinfo {author} {\bibfnamefont {R.}~\bibnamefont {Kumar}},
  \bibinfo {author} {\bibfnamefont {A.}~\bibnamefont {Carvalho}}, \bibinfo
  {author} {\bibfnamefont {K.}~\bibnamefont {{Kumar Amara}}}, \bibinfo {author}
  {\bibfnamefont {W.}~\bibnamefont {Zhao}}, \bibinfo {author} {\bibfnamefont
  {S.}~\bibnamefont {Wang}}, \bibinfo {author} {\bibfnamefont {M.}~\bibnamefont
  {Toh}}, \bibinfo {author} {\bibfnamefont {R.~M.}\ \bibnamefont {Ribeiro}},
  \bibinfo {author} {\bibfnamefont {a.~H.}\ \bibnamefont {{Castro Neto}}},
  \bibinfo {author} {\bibfnamefont {K.}~\bibnamefont {Matsuda}}, \ and\
  \bibinfo {author} {\bibfnamefont {G.}~\bibnamefont {Eda}},\ }\href {\doibase
  10.1038/ncomms5543} {\bibfield  {journal} {\bibinfo  {journal} {Nature
  communications}\ }\textbf {\bibinfo {volume} {5}},\ \bibinfo {pages} {4543}
  (\bibinfo {year} {2014})}\BibitemShut {NoStop}%
\bibitem [{\citenamefont {Kumar}\ and\ \citenamefont
  {Ahluwalia}(2012{\natexlab{b}})}]{Kumar2012b}%
  \BibitemOpen
  \bibfield  {author} {\bibinfo {author} {\bibfnamefont {A.}~\bibnamefont
  {Kumar}}\ and\ \bibinfo {author} {\bibfnamefont {P.~K.}\ \bibnamefont
  {Ahluwalia}},\ }\href {\doibase 10.1140/epjb/e2012-30070-x} {\bibfield
  {journal} {\bibinfo  {journal} {The European Physical Journal B}\ }\textbf
  {\bibinfo {volume} {85}},\ \bibinfo {pages} {186} (\bibinfo {year}
  {2012}{\natexlab{b}})}\BibitemShut {NoStop}%
\end{thebibliography}%

\end{document}